\documentclass[lettersize,journal]{IEEEtran}
\usepackage{amsmath,amsfonts}
\usepackage{amssymb}
\usepackage{algorithmic}
\usepackage{algorithm}
\usepackage{array}
\usepackage[caption=false,font=normalsize,labelfont=sf,textfont=sf]{subfig}
\usepackage{textcomp}
\usepackage{stfloats}
\usepackage{url}
\usepackage{verbatim}
\usepackage{graphicx}
\usepackage{bbding}
\usepackage{orcidlink}
\usepackage{pifont}
\usepackage{wasysym}
\usepackage[]{cite}
\usepackage{enumitem}
\usepackage{titlesec}
\usepackage{tikz}

\hypersetup{
	colorlinks=true,
	linkcolor=black,
	citecolor=black
}
\begin{document}

\title{AEAKA: An Adaptive and Efficient Authentication and Key Agreement Scheme for IoT in Cloud-Edge-Device Collaborative Environments}
\author{Kexian Liu$^{~\orcidlink{0000-0003-3975-5606}}$, Jianfeng Guan$^{~\orcidlink{0000-0002-4411-0741}}$,~\IEEEmembership{Member,~IEEE}, Xiaolong Hu, Jing Zhang, Jianli Liu, Hongke Zhang$^{~\orcidlink{0000-0001-8906-813X}}$,~\IEEEmembership{Fellow,~IEEE}
	\thanks{This work was supported by the National Key R\&D Program of China under Grant No. 2022YFB3102304 and in part by National Natural Science Foundation of China Grants(62394323, 62225105, 62001057) \textit{(Corresponding authors: Jianfeng Guan.)}
		
		Kexian Liu, Jianfeng Guan, Xiaolong Hu, Jing Zhang, and Jianli Liu are with the State Key Laboratory of Networking and Switching Technology, Beijing University of Posts and Telecommunications, Beijing 100876, China (e-mail: kxliu@bupt.edu.cn; jfguan@bupt.edu.cn; hxl814446051@bupt.edu.cn; jzhang2021@bupt.edu.cn; kuohao233@bupt.edu.cn).
		
		Hongke Zhang is with the School of Electronic and Information Engineering, Beijing Jiaotong University, Beijing 100044, China (e-mail: hkzhang@bjtu.edu.cn).
	}
}
\markboth{Journal of \LaTeX\ Class Files,~Vol.~14, No.~8, August~2021}%
{Shell \MakeLowercase{\textit{et al.}}: A Sample Article Using IEEEtran.cls for IEEE Journals}
\maketitle

\markboth{Journal of \LaTeX\ Class Files,~Vol.~14, No.~8, August~2021}%
{Shell \MakeLowercase{\textit{et al.}}: A Sample Article Using IEEEtran.cls for IEEE Journals}


\maketitle

\begin{abstract}
To meet the diverse needs of users, the rapid advancement of cloud-edge-device collaboration has become a standard practice. However, this complex environment, particularly in untrusted (non-collaborative) scenarios, presents numerous security challenges. Authentication acts as the first line of defense and is fundamental to addressing these issues. Although many authentication and key agreement schemes exist, they often face limitations such as being tailored to overly specific scenarios—where devices authenticate solely with either the edge or the cloud—or being unsuitable for resource-constrained devices. To address these challenges, we propose an adaptive and efficient authentication and key agreement scheme (AEAKA) for Cloud-Edge-Device IoT environments. This scheme is highly adaptive and scalable, capable of automatically and dynamically initiating different authentication methods based on device requirements. Additionally, it employs an edge-assisted authentication approach to reduce the load on third-party trust authorities. Furthermore, we introduce a hash-based algorithm for the authentication protocol, ensuring a lightweight method suitable for a wide range of resource-constrained devices while maintaining security. AEAKA ensures that entities use associated authentication credentials, enhancing the privacy of the authentication process. Security proofs and performance analyses demonstrate that AEAKA outperforms other methods in terms of security and authentication efficiency.
\end{abstract}

\begin{IEEEkeywords}
Cloud-Edge-Device, Internet of Things (IoT), Authentication and Key Agreement, Session Key.
\end{IEEEkeywords}

\section{Introduction}
\IEEEPARstart{W}{ith} the rapid advancement of information technology, cloud computing has become essential for meeting diverse user needs. Its applications range from online medical diagnostics to dashboard camera video transmission and daily conversations\cite{he2023taxonomy}. While cloud computing provides various services, it has limitations for time-sensitive applications, such as real-time data processing and low-latency communication\cite{li2021user,yang2023burst}. Edge computing addresses these challenges by relocating computation and data storage to the network edge, reducing latency and enhancing real-time capabilities\cite{zhou2022service}. However, edge computing's computational and storage capabilities are limited, requiring cloud support for large-scale tasks\cite{yao2022blockchain,yang2023detfed}. Current research \cite{wu2023survey,hu2023intelligent,zhang2020efficient,lin2023framework} focuses on optimizing the cloud-edge-device architecture, combining the strengths of cloud and edge computing to meet diverse user needs, making it a trend in Internet of Things (IoT) development. A three-layer Cloud-Edge-Device architecture is shown in Fig. 1.\par

In this intricate environment, security is a paramount concern. In April 2024, SOCRadar's researchers uncovered a Microsoft data breach, exposing employee credentials and internal Bing search engine documents on a public Azure cloud server. Despite the vulnerability being addressed, there's a risk that malicious actors accessed the data, jeopardizing service security. Similarly, BoAt Lifestyle recently leaked sensitive information of approximately 7.5 million customers, including names, addresses, and contact details. These incidents emphasize the urgent need for robust authentication measures among devices, edge servers, and cloud servers to prevent unauthorized access to sensitive data. Addressing this imperative, an effective authentication and key agreement scheme, facilitating mutual authentication and session key generation, is essential to bolster protection against various security threats.\par

\begin{figure}[!t]
	\centering
	\includegraphics[width=0.5\textwidth]{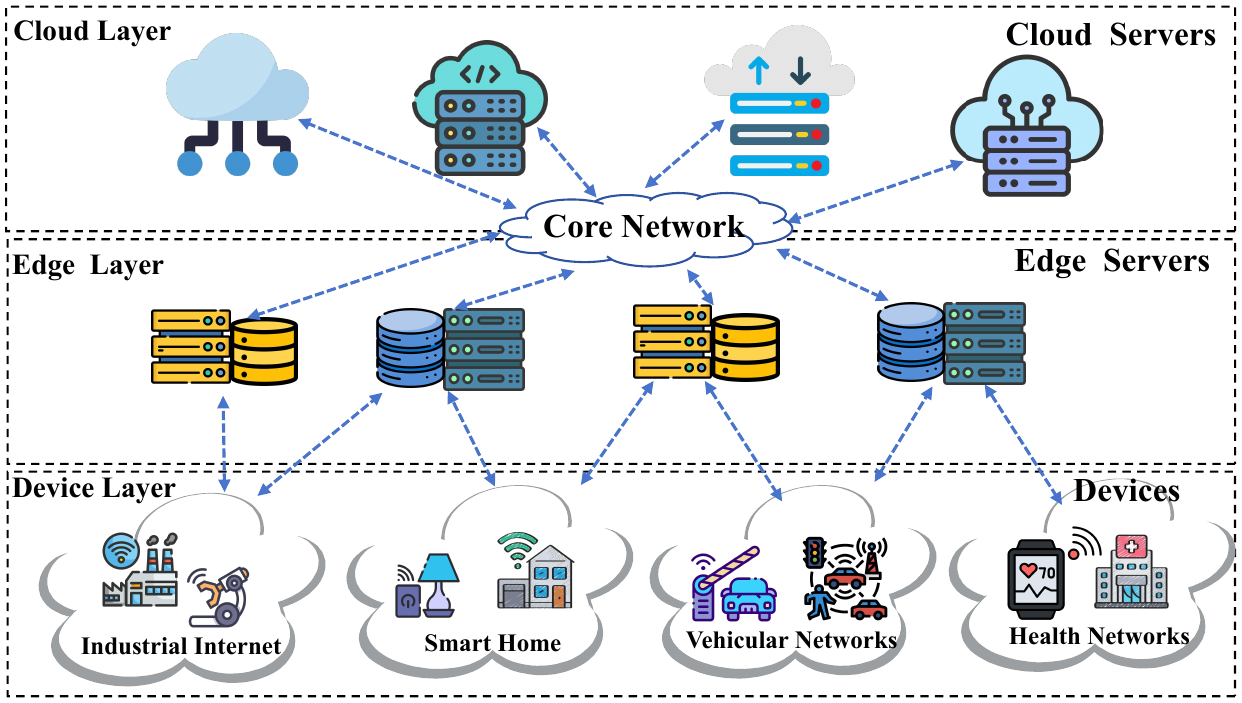}
	\caption{Cloud-Edge-Device architecture overview.}
	\label{fig_1}
	\vspace{-1.6em}
\end{figure}
Currently, existing authentication and key agreement schemes in cloud-edge-device environments, such as device-to-cloud\cite{cui2019extensible,zhang2020smaka} and device-to-edge schemes\cite{xie2024effectively,wei2021lightweight}. However, these schemes typically apply to single scenarios and do not consider comprehensive cloud-edge-device collaborative scenarios. Due to the complexity of the cloud-edge-device collaborative environment, where time-sensitive tasks may require edge computing and tasks with high computational and storage demands may require cloud computing\cite{yao2022blockchain,yang2023detfed,yang2023burst}. Therefore, there is a need for an  adaptive, scalable, and highly available cloud-edge-device collaborative authentication scheme.\par
It is evident that combining these single-scenario schemes into a unified cloud-edge-device framework is impractical because they have different system initialization requirements and employ different encryption algorithms. Furthermore, these schemes also have other drawbacks. For instance, in schemes \cite{cui2019extensible,zhang2020smaka,wei2021lightweight,mahmood2022seamless}, the frequent interaction with trust authorities(TAs) or registration centers(RCs) during the authentication process can lead to an excessive load on TAs and RCs, potentially causing security issues\cite{almarshoud2024security,zhang2019pa}. Additionally, in schemes \cite{zhang2024blockchain,lee2023secure,lyu2022a2ua}, the use of complex Elliptic Curve Cryptography (ECC) operations and bilinear pairings during authentication may not be suitable for resource-constrained devices, increasing the difficulty of deployment.

This research aims to design an adaptive and highly available cloud-edge-device collaborative authentication and key agreement scheme to address the shortcomings of current schemes. Authentication and key agreement in this collaborative environment faces several major challenges: 
\begin{itemize}
	\item[$\bullet$] How to design a highly adaptive and scalable cloud-edge-device collaborative authentication and key agreement architecture? As mentioned earlier, existing research typically considers single authentication scenarios.
	
	\item[$\bullet$] How to design an authentication and key agreement protocol that does not rely on third-party TAs, to reduce the load and burden on TAs? 
	
	\item[$\bullet$] How to design a universally applicable authentication and key agreement protocol for diverse resource-constrained devices? Some complex cryptographic algorithms, such as ECC and bilinear pairings, are unsuitable for resource-constrained devices.
	
	\item[$\bullet$] How to achieve authentication and key agreement without disclosing critical device privacy? To address issues related to insider privilege attacks or semi-trusted servers, devices should complete the authentication process without disclosing their own privacy.
\end{itemize}\par

To address these challenges, we propose the following solutions: First, design a highly available cloud-edge-device collaborative authentication architecture that adaptively initiates different authentication methods according to the specific needs of devices, reducing the complexity of device operations. Second, design an edge-assisted authentication scheme that does not frequently rely on TAs, thereby reducing the load on TAs. Third, We propose a hash-based authentication protocol, making it suitable for most resource-constrained devices. Finally, entities generate related authentication credentials during registration, which are used during subsequent authentication processes, preventing the disclosure of device privacy.
The main contributions of this paper include the following:
\begin{itemize}

\item[$\bullet$] Designing an adaptive and efficient authentication and key agreement architecture in cloud-edge-device collaborative environments. Devices first request the edge server. If the edge server can meet the service demand, only authenticaiton between the device and the edge server is required. If the edge server cannot meet the service demand, it automatically conducts mutual authentication among the device, the edge server, and the cloud server. The architecture adaptively initiates different authentication and key agreement methods during service requests, reducing device operations' complexity and enhancing the efficiency of the authentication process and response speed.

\item[$\bullet$] Proposing an edge-assisted authentication and key agreement sheme. In this architecture, the authentication and key agreement process no longer frequently relies on third-party TAs. With the assistance of the edge server, mutual anonymous authentication between devices and edge servers and between devices and cloud servers can be achieved. Reducing reliance on TAs not only alleviates the burden on TAs but also mitigates the security risks associated with frequent TAs computations.

\item[$\bullet$] Proposing a lightweight authentication and key agreement method. This method uses a hash-based algorithm during the authentication process, avoiding complex ECC operations, thereby significantly reducing the computational overhead of authentication. The lightweight authentication method makes this scheme particularly suitable for various resource-constrained IoT devices, ensuring that even devices with limited performance can perform efficient authentication operations, expanding the overall system's applicability and practicality.

\item[$\bullet$] Introducing associated authentication credentials in the authentication and key agreement protocol. Entities generate associated authentication credentials during registration, which are used during subsequent authentication processes. By pre-generating and using associated credentials, the efficiency and security of the authentication process are ensured.

\item[$\bullet$] Formal security proof and performance analysis. We formally prove the security of AEAKA and analyze its computational and communication overhead. The results demonstrate that AEAKA offers high security and performance.
	
\end{itemize}\par

The remainder of the paper is organized as follows. In Section II, we review the existing authentication and key agreement mechanisms in cloud-edge-device environment. Section III provides system model of AEAKA. In Section IV, we elaborate on AEAKA. Section V conducts a security performance analysis of AEAKA. In Section VI, we establish an experimental environment to evaluate the performance of the proposed scheme and compare it with other alternatives. Finally, Section VII concludes the paper.

\section{Related work}
\renewcommand\arraystretch{1.4}
\begin{table*}[h]
	\caption{SUMMARIZING EXISTING SCHEMES}
	\centering  
	\begin{tabular}{cccc}
		\hline
		\textbf{Reference} & \textbf{Context} &\textbf{Operations} &\textbf{Entities involved in authentication}\\ 
		\hline
		Cui \textit{et al.}\cite{cui2019extensible}&Cloud Computing & ECC,XOR,Hash& Vehicle,Cloud Service Provider,TA\\ 
		Zhang \textit{et al.}\cite{zhang2020smaka}&Cloud Computing & ECC,XOR,Hash,Sym.En/De,& Vehicle,Cloud Service Provider,TA\\
		Xie \textit{et al.}\cite{xie2024effectively}& Edge Computing& XOR,Hash& TE,Authentication Server,Computing Server\\ 
		Wei \textit{et al.} \cite{wei2021lightweight}& Edge(Fog) Computing&ECC,XOR,Hash,Lagrange interpolation & Vehicle,Fog Server,TA\\ 
		Mahmood \textit{et al.}\cite{mahmood2022seamless} &Edge Computing &ECC,XOR,Hash & Mobile device, ES, RC\\ 
		Zhang \textit{et al.}\cite{zhang2024blockchain} & Edge Computing &ECC,XOR,Hash, Bilinear pairing & Device,ES,TA\\ 
		Lee \textit{et al.}\cite{lee2023secure}&Edge Computing &ECC,XOR,Hash,Bilinear pairing,Sym.En/De & User,ES\\ 
		Lyu \textit{et al.}\cite{lyu2022a2ua}&Cloud Computing & ECC,XOR,Hash,Bilinear pairing& User,Cloud Service Provider\\ 
		Xu \textit{et al.}\cite{xu2022efficient}&Edge Computing &ECC,XOR,Hash & User,ES\\ 
		Mahmood \textit{et al.}\cite{mahmood2023design}&Edge Computing &ECC,XOR,Hash,Sym.En/De & User,ES,CS\\ 
		Vangala \textit{et al.}\cite{vangala2022blockchain}& Edge(Fog) Computing&ECC,XOR,Hash & Device,Mobile Vehicle,Fog Server\\ 
		Ma \textit{et al.}\cite{ ma2023stcla}&Edge(Fog) Computing & ECC,XOR,Hash& Vehicle,Fog Server\\ 
		Badshah \textit{et al.}\cite{badshah2022aake}&Cloud Computing & ECC,XOR,Hash& Vehicle,RSU,CS\\ 
		Ren \textit{et al.}\cite{ren2023provable}&Cloud Computing & ECC,XOR,Hash& Device,CS\\ 
		Kumar \textit{et al.}\cite{kumar2020lightweight}&Cloud Computing & ECC,XOR,Hash& Device,CS\\ 
		Liu \textit{et al.}\cite{liu2022lightweight}&Cloud Computing & ECC,XOR,Hash,Bilinear pairing& User,CS\\ 
		\hline
	\end{tabular}
\end{table*}

In this section, we introduce exiting authentication and key agreement schemes in the cloud-edge-device environment, which can be mainly divided into two categories: device-to-edge and device-to-cloud. Table I provides a detailed analysis of various existing authentication and key agreement schemes in different contexts, including operations and entities involved in authentication. Here, ES represents the Edge Server, CS represents the Cloud Server, and Sym.En/De indicates symmetric encryption/decryption.

\subsection{Device-to-Edge Authentication and Key Agreement}
There are several schemes in the field of mobile edge computing.\par
Mahmood \textit{et al.}\cite{mahmood2022seamless} proposed an identity authentication protocol suitable for moblie edge computing (MEC) environments. This protocol involves mobile users and MEC, using complex ECC algorithms to achieve secure and efficient communication between all target entities while maintaining user anonymity. However, this scheme has some issues, such as the frequent interaction with the RC, which can lead to high complexity at the RC. Zhang \textit{et al.}\cite{zhang2024blockchain} proposed an authentication framework for IoT environments, which establishes secure communication between devices and edge servers. However, the scheme has drawbacks, such as the use of complex ECC and bilinear pairing algorithms, and the need to request third-party entities for entity authentication, leading to decreased efficiency. Xie \textit{et al.}\cite{xie2024effectively} proposed a new anonymous authentication key agreement scheme that considers the resource constraints of terminal devices and the security risks of semi-trusted servers. However, it only implements authentication between terminal equipment (TE) and ES by setting up an authentication proxy server (AS). Xu \textit{et al.}\cite{xu2022efficient} created an efficient identity authentication protocol with provable security and anonymity for MEC, which ensures user anonymity during communication. However, this scheme uses ECC algorithms, which are unsuitable for resource-constrained devices, and it only achieves authentication between users and MEC, not applicable to the cloud-edge-device scenario. Mahmood \textit{et al.}\cite{mahmood2023design} developed a key agreement solution for mobile users to achieve single-round mutual authentication. However, it only implements authentication between users and ES. Lee \textit{et al.}\cite{lee2023secure} proposed a secure and anonymous authentication scheme that without a trusted third party. However, the scheme uses complex ECC and bilinear pairing operations, which are not user-friendly, and it only implements authentication between users and ES.\par

Additionally, there are also schemes in the field of fog computing(FN).\par
Wei \textit{et al.} \cite{wei2021lightweight}proposed a lightweight, conditional privacy-preserving authenticated key agreement scheme, which uses symmetric cryptography methods to reduce the overhead of the authenticated key agreement process. Additionally, a multi-TA model was considered to solve the single-point failure problem. Vangala \textit{et al.}\cite{vangala2022blockchain} designed an efficient blockchain-based authentication key agreement scheme, AgroMobi Block, which achieving authentication between devices, mobile vehicles, and FN. Ma \textit{et al.}\cite{ ma2023stcla} designed an efficient authentication key agreement scheme (STCLA) for fog-based vehicular networks. Similarly, the complex ECC operations are not user-friendly, and it only implements authentication between users and FN.

\subsection{Device-to-Cloud Authentication and Key Agreement}
Badshah \textit{et al.}\cite{badshah2022aake} designed an anonymous authenticated key exchange mechanism AAKE-BIVT for blockchain-enabled intelligent transportation. Vehicles authenticate with nearby roadside units (RSUs), which then authenticate with the CS, securely aggregating data from related RSUs and generating transactions. Tu \textit{et al.} \cite{tu2023eake} proposed an efficient anonymous authenticated key exchange scheme (EAKE-WC) for wearable computing. This scheme facilitates mutual authentication between users and wearable devices, as well as between cloud servers and users. Cui \textit{et al.}\cite{cui2019extensible} proposed a robust and extensible authentication scheme for vehicular networks, which vehicles only need to register once with a Trusted Authority (TA) to achieve quick and efficient authentication with Cloud Service Providers (CSPs). Based on Cui \textit{et al.}\cite{cui2019extensible}'s scheme, Zhang \textit{et al.}\cite{zhang2020smaka} proposed a many-to-many authentication and key agreement scheme for secure authentication between multiple vehicles and CSPs. When a vehicle wants to request services from multiple CSPs, it only needs to send one request message, significantly reducing computational and communication overhead. Ren \textit{et al.}\cite{ren2023provable} proposed an anonymous certificateless lightweight identity authentication protocol (ACLAP) based on ECC. This scheme uses the device user’s passwords and biometrics as authentication credentials, without storing any trusted evidence on the cloud server, thereby solving the resource consumption issue caused by the large number of devices in the IoT environment. Kumar \textit{et al.}\cite{kumar2020lightweight} proposed an identity-based anonymous authentication and key agreement (IBAAAKA) protocol for cloud-assisted environments in wireless body area networks (WBANs). This protocol achieves mutual authentication between wearable devices and the cloud while ensuring user anonymity. Lyu \textit{et al.}\cite{lyu2022a2ua} proposed an auditable anonymous user authentication (A2UA) protocol based on blockchain for cloud services. The A2UA protocol primarily uses bilinear pairings, partial authentication factors, dynamic credits, and fake public keys to achieve anonymous mutual authentication between users and cloud service providers. Liu \textit{et al.}\cite{liu2022lightweight} proposed a lightweight and secure editable signature scheme with coarse-grained additional redaction control (CRS) for secure dissemination of healthcare data in cloud-assisted healthcare IoT systems. This scheme achieves authentication between users and the cloud, ensuring secure propagation of healthcare data.

In conclusion, existing research primarily focuses on single authentication scenarios, neglecting the collaborative authentication needs of cloud-edge-device environments. Additionally, many schemes overly rely on TA or RC during the authentication process, which can increase the burden on TAs and RCs and potentially lead to security issues. Furthermore, many schemes are based on complex ECC algorithms, which are not suitable for resource-constrained devices.

Therefore, designing a highly scalable and adaptive cloud-edge-device collaborative authentication architecture is essential. This scheme should avoid reliance on TAs during authentication and adopt lightweight authentication methods suitable for resource-constrained IoT devices, ensuring high practicality and security. We believe that this innovative authentication scheme will provide robust support and security assurance for future cloud-edge-device collaborative computing.

\section{SYSTEM MODEL}

In this section, we present the proposed system model and certain assumptions underlying the system, culminating in the delineation of our security objectives and threat model.
\subsection{Network Model and Assumptions}
\begin{figure}[!t]
	\centering
	\includegraphics[width=0.5\textwidth]{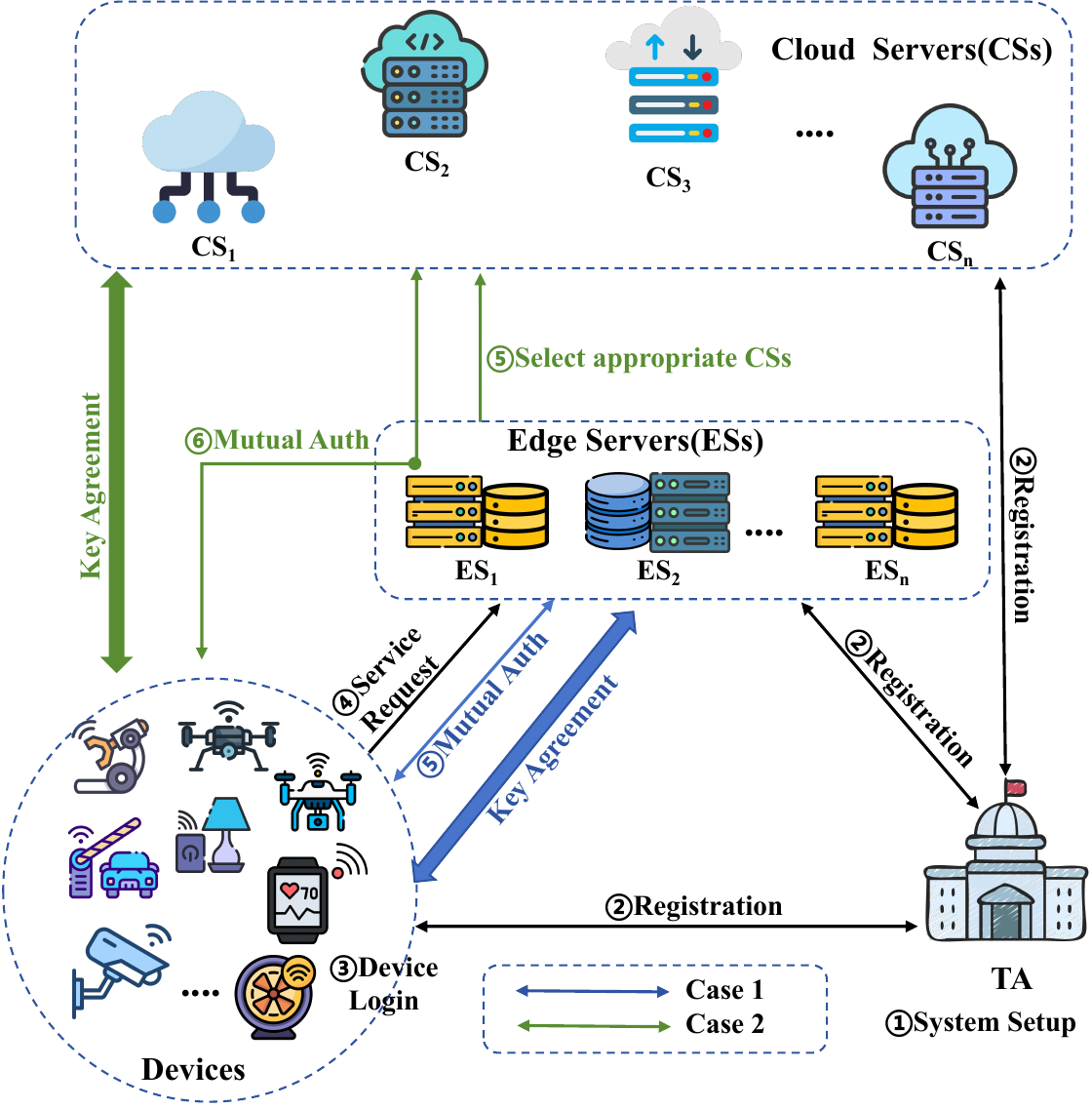}
	\caption{AEAKA system model.}
	\label{fig_1}
	\vspace{-1.6em}
\end{figure}
Fig. 2 illustrates the system model of the cloud-edge-device authentication system proposed herein, comprising a Trusted Authority (TA), numerous devices, multiple Edge Servers (ESs), and multiple Cloud Servers (CSs). The functionalities and system assumptions of these entities are elaborated below.

{\bf{\textit{1) Trusted Authority (TA)}}}: This entity stands as a reliable, widely deployed, and highly secure entity entrusted with the operational oversight of governmental intelligent transportation systems. Endowed with tamper-resistance capabilities alongside ample storage and computational prowess, the TA assumes the role of a registrar for devices, ESs, and CSs, dispensing trusted credentials to each entity while retaining the capability to trace their authentic identities. In this discourse, to forestall single points of failure, the TA operates as a distributed collective\cite{liu2024dkgauth}, bolstered by sundry consistency algorithm protocols to ensure perpetually veritable trustworthiness. It is postulated to perpetually maintain authenticity and integrity, impervious to compromise.

{\bf{\textit{2) Cloud Server (CS)}}}: Cloud Server can be regarded as Cloud Service Provider(CSP). CSs furnish diverse network services to devices, with each CS potentially offering distinct services. To augment the multifunctional utility of devices, CSs periodically broadcast security and entertainment-related services to TA or edge servers. Furthermore, CSs can offload computational tasks from devices and ESs, shouldering burdensome computations.

{\bf{\textit{3)  Edge Server (ES)}}}: Capable of communication with other entities such as devices, CSs, and TA, each ES is confined to communication within its geographical ambit. Upon the ingress of authenticated devices into an ES's geographical domain, it undertakes the dissemination of pertinent authentication-related public messages. Within the communication purview of an ES, concurrent authentication and communication with multiple devices are plausible. The computational and storage capacities of ESs are presumed to be inferior to CSs. Additionally, ESs serve to facilitate authentication between devices and CSs, aiding in the optimal selection of CS clusters to furnish devices with premium services. Finally, ESs engage in communication with TA during the device authentication process to ascertain the genuine identities of suspicious devices.

{\bf{\textit{4) Device}}}: Endowed with commendable wireless communication capabilities albeit constrained computational resources and storage space, devices serve as both the instigators and beneficiaries of network service requests.

The elucidated system model and attendant assumptions serve as the bedrock upon which our subsequent discourse concerning security objectives and threat models is predicated.

\subsection{Security Objectives}

\begin{itemize}
	\item[$\bullet$] {\bf{\textit{Anonymity}}}: To mitigate the risk of privacy breaches, devices adopt anonymous identities during authentication, refraining from divulging any secrets to relevant servers.
\end{itemize}

\begin{itemize}
	\item[$\bullet$] {\bf{\textit{Unlinkability}}}: Devices can randomly select pre-generated $pid^{(x)}_{i,j}$ during each authentication, thereby imbuing the process with unlinkability characteristics. And no third party can link intercepted messages to the same device.
\end{itemize}

\begin{itemize}
	\item[$\bullet$] {\bf{\textit{Mutual Authentication}}}: Reciprocal verification between devices, ESs, and CSs is imperative to safeguard the system against impersonation attacks.
\end{itemize}

\begin{itemize}
	\item[$\bullet$] {\bf{\textit{Anonymous Session Key Agreement}}}: Anonymous session key agreement is imperative to ensure the confidentiality of information pertaining to entertainment and security within the network.
\end{itemize}

\begin{itemize}
	\item[$\bullet$] {\bf{\textit{Resilience Against Common Attacks}}}: The proposed scheme must demonstrate resilience against prevalent attack vectors such as replay attacks, offline password guessing attacks, and impersonation attacks, thereby ensuring the overall security of the vehicular network.
\end{itemize}

\subsection{Threat Model}

In the proposed framework, we adopt the well-known Dolev-Yao threat model \cite{dolev1983security}. According to this model, adversary $\mathcal{A}$ possesses the capability to eavesdrop on, modify, delete, forge, replay, or even inject false information into communications traversing insecure public channels between two communicating parties. Devices, ESs, and CSs are deemed insecure, as any entity transmits information over public channels. Furthermore, adversaries can emulate the behaviors of communication entities, TA is considered entirely trustworthy, impervious to adversary compromise, with no information stored on TA susceptible to leakage to adversaries.

\section{OUR PROPOSED SCHEME}

\renewcommand\arraystretch{1.4}
\begin{table}[h]
	\caption{Summary of abbreviations and notation}
	\centering  
	\begin{tabular}{p{2.4cm}p{5.6cm}}
		\hline
		\textbf{Notation} & \textbf{Description}\\ 
		\hline
		$TA$ & Trusted authority \\
		$User_i$ & i-th user \\
		$D_i,ES_j,CS_k$ & i-th device, j-th ES and  k-th CS, respectively \\
		$ID_i,EID_j,CID_k$ & Real identities of $D_i,ES_j,CS_k$, respectively\\
		$UID_i$ & Real identity of $User_i$ \\
		$DID_i$ &  Identitier of $D_i$ and $User_i$\\
		$PW_i$ &   Password of $User_i$\\
		$pid^{(x)}_{i,j}$ & x-th pseudonym of entity $i$ to entity $j$.\\
		$PID_{i,j}$ & Set of pseudonym of entity $i$ to entity $j$.\\
		$a^{(x)}_{i,j}, b^{(x)}_{i,j}$ & $D_i$'s x-th secret value to $ES_j$.\\
		$B_{i,j}$ & Set of secret value of entity $i$ to entity $j$. \\
		$s$ & $TA$'s pivate key.\\
		$SE_j,SC_k$ & Secret values of $ES_j,CPS_k$, respectively\\
		$M_i$ & Messages of entities.\\
		$\alpha,\beta,\theta,\nu,EPW_i$ & Hash values\\
		$x_i, A_{x,y}, S_{x,y}$ & Temporary Values\\
	    $sk_{ij}$  & Sessionkey of entity $i$ and $j$\\
	    ${SK_i,PK_i}$ & Private key and public key of entity $i$ \\
		$T_i$ & Timestamp\\
		$h$ & Collision-resistant hash function\\
		$||$ & Concatenation operation\\
		$\oplus$ & XOR operation\\
		\hline
	\end{tabular}
\end{table}

In this section, we provide a detailed description of AEAKA. Table II enumerates the primary symbols used throughout the five phases of this study along with their corresponding definitions. The first stage involves system initialization, where TA allocates public parameters to the system. The second stage encompasses the legitimate registration of users, devices, ESs, and CSs with TA to acquire the requisite authentication credentials. The third stage mandates users to undergo necessary device login to proceed further. The fourth stage entails authentication and key agreement before device requests for various services. This stage bifurcates into two scenarios: the first involves mutual authentication and key agreement between devices and ESs, while the second entails, with ESs' assistance, mutual authentication and key agreement between devices and multiple CSs. The fifth stage pertains to user password updates. The scheme exhibits the following advantages:  1) AEAKA method demonstrates heightened scalability, catering more effectively to diverse service requisites of devices, with CS registration and selection remaining transparent to users and known only to ESs. 2) Authentication and key agreement processes do not necessitate TA involvement, thus substantially reducing TA's communication overhead. 3) AEAKA eschews complex ECC operations during authentication and key agreement, thereby enhancing system efficiency.

\subsection{Setup Process}

At this phase, TA initializes the system with public parameters based on our previous work, DKGAuth \cite{liu2024dkgauth}. These parameters are then disseminated publicly within the network. Each entity within the network can compute the requisite authentication materials based on these public parameters. \par
Primarily, TA selects a secret value $s$ as its private key, known to all distributed TA nodes. TA also selects a secure, collision-resistant, one-way hash function $h: \{0, 1\}^* \to \{0, 1\}^l$, here $l$ denotes the limit length of bit string\cite{sarkar2010simple,stinson2006some}. \par
It's worth noting that ECC is presented as an optional component. While the proposed scheme's authentication process does not involve ECC operations, ECC may be used in the future to establish secure channels between TA nodes, ESs, and CSs, among other applications.

\subsection{Registration Process}

At this phase, devices, EDs, and CSs respectively provide registration information to TA via secure channels for registration.

$\bullet$ {\bf{CSs Registration}}

Fig. 3 illustrates the process of $CS_k$ registration with TA through a secure channel.\par

\begin{itemize}
	\item[1)] $CS_k$ first selects its identifier $CID_k$ and then submits its registration request $\{CID_k\}$ to TA through the secure channel.
	
	\item[2)] Upon receiving the registration request, TA first checks the legality of $CID_k$, such as checking for duplicate registrations. If $CID_k$ is deemed illegal, the request is disregarded. Otherwise, TA employs the DKGAuth($CID_k$) algorithm\cite{liu2024dkgauth} to generate a long-term public-private key pair $(SK_k, PK_k)$ for $CS_k$. It's worth noting that $(SK_k, PK_k)$ is strongly bound to $CID_k$ for future $CS_k$ use cases. Subsequently, TA computes the authentication credential $SC_k = h(s || h(PK_k))$ for $CS_k$.
	
	\item[3)] TA returns the $\{SC_k, SK_k\}$ message to $CS_k$. Finally, $CS_k$ stores $\{SC_k,SK_k\}$, TA stores $\{CID_k, PK_k\}$ in $ListCS()$.

\end{itemize}
\begin{figure}[!t]
	\centering
	\includegraphics[width=3.6in]{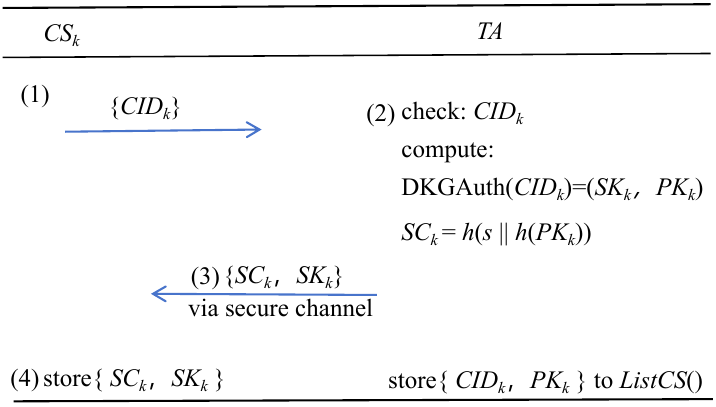}
	\caption{CS registration.}
	\label{fig_1}
\end{figure}

It should be noted that when registering, the cloud server can provide TA with the services it offers, allowing ES to assist users in choosing the most appropriate services. Since this paper focuses on the authentication part, the registration of services is not within the scope of this discussion and will not be elaborated on further.

$\bullet$ {\bf{ESs Registration}}

Fig. 4 illustrates the process of $ES_j$ registration with TA through a secure channel.\par
\begin{itemize}
	\item[1)] $ES_j$ first selects its identifier $EID_j$ and then submits its registration request $\{EID_j\}$ to TA via a secure channel.
	
	\item[2)] Upon receiving the registration request, TA first checks the legality of $EID_j$, such as checking for duplicate registrations. If $EID_j$ is deemed illegal, the request is disregarded. Otherwise, TA employs the DKGAuth($EID_j$) algorithm to generate a long-term public-private key pair $(SK_j, PK_j)$ for for $ES_j$. Then, TA generates the authentication credential $SE_j = H(s || h(PK_j))$, where $SE_j$ is used for validating device identity. To ensure anonymity in the authentication between $ES_j$ and $CS_k$, TA computes pseudonyms $pid_{j,k} = h(EID_j || h(PK_k))$ and authentication credentials $C_{j,k} = h(pid_{j,k} || h(s || h(PK_k)))$ for $ES_j$, where $PK_k$ represents the public key of $CS_k$. Finally, TA returns the response message $\{SE_j, SK_j ,pid_{j,k}, C_{j,k}\}$ to $ES_j$.
	
	\item[3)] Upon receiving the response message from TA, $ES_j$ stores $\{SE_j,SK_j ,(pid_{j,k}, C_{j,k})\}$ to $ListE2C()$ while TA saves $\{EID_j,(pid_{j,k}, PK_j)\}$ to $ListES()$.
	
\end{itemize}

\begin{figure}[!t]
	\centering
	\includegraphics[width=3.6in]{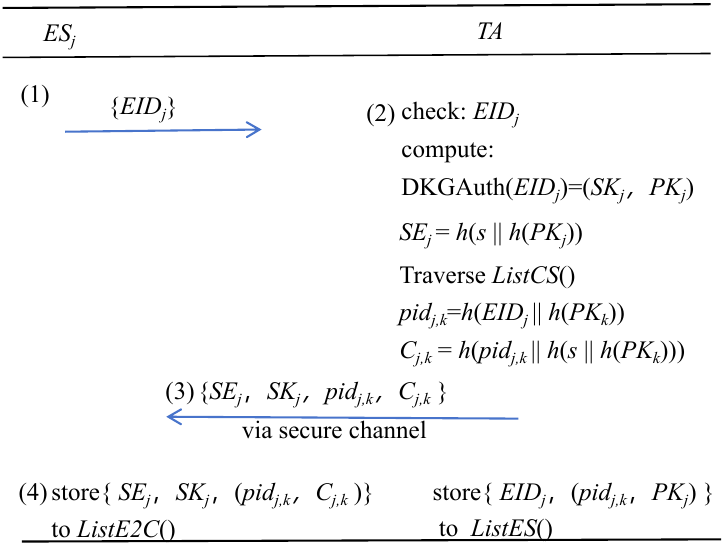}
	\caption{ES registration.}
	\label{fig_1}
\end{figure}

It's worth noting that the $(pid_{j,k}, C_{j,k})$ generated in the registration process serve as the anonymous identity and authentication credentials for communication between $ES_j$ and $CS_k$. In actual scenarios, when $ES_j$ registers, TA customizes a sequence of CSs based on preferences and geographical location provided by $ES_j$, selecting multiple CSs from the registered list, generating multiple $(pid_{j,x}, C_{j,x})$ pairs for $ES_j$. This enables $ES_j$ to provide more services for devices within its scope.

$\bullet$ {\bf{Devices Registration}}

Fig. 5 illustrates the process of $User_i$ and $D_i$ registration with TA via a secure channel.\par

\begin{itemize}
	\item[1)] Firstly, the device user $User_i$ selects username $UID_i$ and password $PW_i$, then computes $EPW_i$ = $h(UID_i || PW_i)$, and finally submits $\{UID_i, ID_i, EPW_i\}$ to TA, where $ID_i$ represents the identifier of $D_i$.
	
	\item[2)] Upon receiving the registration request, TA first computes $DID_i = h(UID_i || ID_i || s)$, then checks if $DID_i$ is legal, such as checking for duplicate registrations. If $DID_i$ is deemed illegal, the request is disregarded. Otherwise, TA, if necessary, employs the DKGAuth($DID_i$) algorithm to generate a long-term public-private key pair $(SK_i, PK_i)$ for $DID_i$. To ensure user authentication anonymity and unlinkability, TA generates multiple pseudonyms for the $DID_i$, $pid^{(x)}_{i,j} = h(DID_i || h(PK_j) || T_x)$, where $Tx$ represents the timestamp. Then computes authentication credentials $a^{(x)}_{i,j} = h(pid^{(x)}_{i,j} || h(s || h(PK_j)))$ and $b^{(x)}_{i,j} = EPW_i \oplus a^{(x)}_{i,j}$, where $PK_j$ represents the public key of $ES_j$.  Finally, TA returns $\{DID_i,SK_i,(h(PK_j), PID_{i,j}, B_{i,j})\}$ to $D_i$ and stores $\{DID_i, PK_i, PID_{i,j}\}$ to $ListDID()$, where $PID_{i,j} = \{pid^{(1)}_{i,j}, pid^{(2)}_{i,j}, ..., pid^{(n)}_{i,j}\}$, $B_{i,j} = \{b^{(1)}_{i,j}, b^{(2)}_{i,j}, ..., b^{(n)}_{i,j}\}$. It's noteworthy that during user device registration, apart from $ES_j$, authentication credentials can also be simultaneously generated with multiple other ESs.
	
	\item[3)] Upon receiving the registration reply from TA, $D_i$ computes $Q_i = h(UID_i || ID_i || PW_i)$, which is used for user $UID_i$ to login to $D_i$. Finally, $D_i$ stores $\{Q_i, DID_i, SK_i, (h(PK_j), PID_{i,j}, B_{i,j})\}$ to $D2EList()$.
	
\end{itemize}
\begin{figure}[!t]
	\centering
	\includegraphics[width=3.6in]{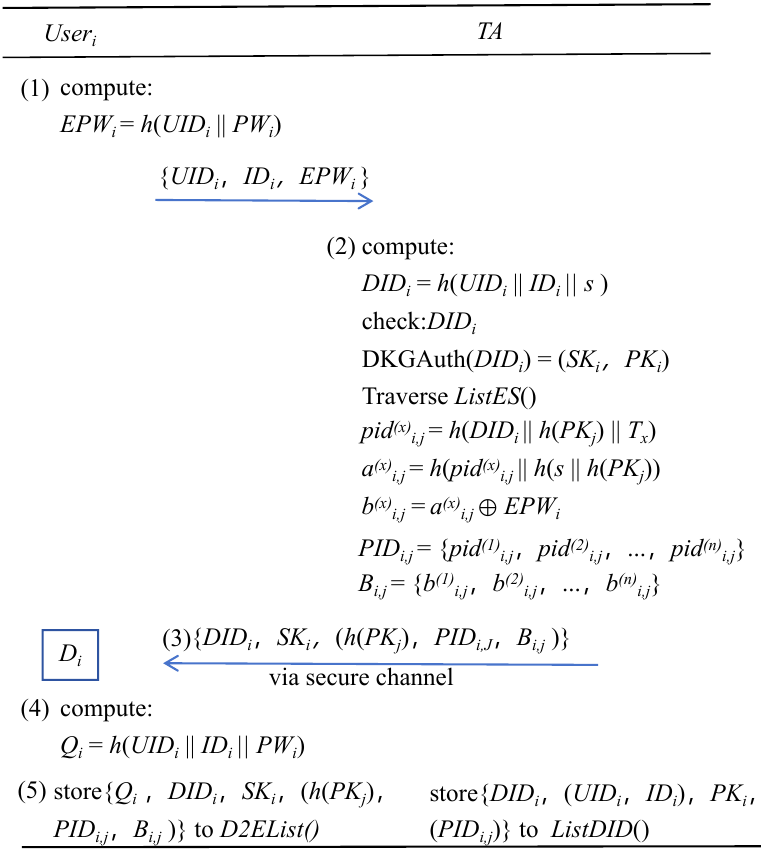}
	\caption{User and device registration.}
	\label{fig_1}
\end{figure}

\subsection{User Login}

The login process serves as a prerequisite for authentication and key agreement. Below is the description of the process wherein $User_i$ logs into device $D_i$:
\begin{itemize}
	\item[1)] Initially, the user $User_i$ inputs $\{UID_i, ID_j, PW_i\}$ to $D_i$.
	
	\item[2)] $D_i$ computes $Q_i' = h(UID_i||ID_j||PW_i)$, then verifies if $Q_i'$ is identical to the pre-stored $Q_i$. If $Q_i'$ matches $Q_i$, the login is deemed successful. Conversely, if $Q_i'$ differs from $Q_i$, the login request will be rejected.

\end{itemize}

\subsection{Authentication and Key Agreement}

After the user $User_i$ successfully logs into device $D_i$, $D_i$ sends its request message to the $ES_j$ within its geographical range instead of broadcasting the message to CSs. Subsequently, $ES_j$ authenticates $D_i'$s legitimacy, and if $D_i$ is legitimate, the $ES_j$ provides the requested service accordingly. To enhance the responsiveness of user requests, two cases are defined. \par

{\bf{{Case 1.}}} This case involves direct authentication and key agreement between $D_i$ and $ES_j$ if the $ES_j$ itself has the capability to fulfill the user's service requirements.

{\bf{{Case 2.}}} The second mode occurs when the $ES_j$ cannot meet the user's service demands, in which case the $ES_j$ recommends the best CS service to the user, which may entail one or multiple CSs. Subsequently, authentication and session key agreement between $D_i$ and CSs are facilitated by the $ES_j$. It's important to note that the selection of suitable CSs lies beyond the scope of this paper. For more details, please refer to \cite{singh2017compliance,gireesha2020iivifs,gireesha2022fuzzy,hussain2020novel,hussain2022cloud,guo2023trust}.

The process of anonymous mutual authentication and session key agreement among $D_i, ES_j$, and $CS_k$ is delineated in detail in Fig. 6. Next, we will provide a detailed description of the process.

$\bullet$ {\bf{\textit{Case 1}}}:

In the scenario where $ES_j$ can fulfill the user's requirements independently, $ES_j$ does not need to request CSs. Therefore, the authentication and key agreement process between $D_i$ and $ES_j$ suffices.

\begin{figure*}[!t]
	\centering
	\includegraphics[width=1\textwidth]{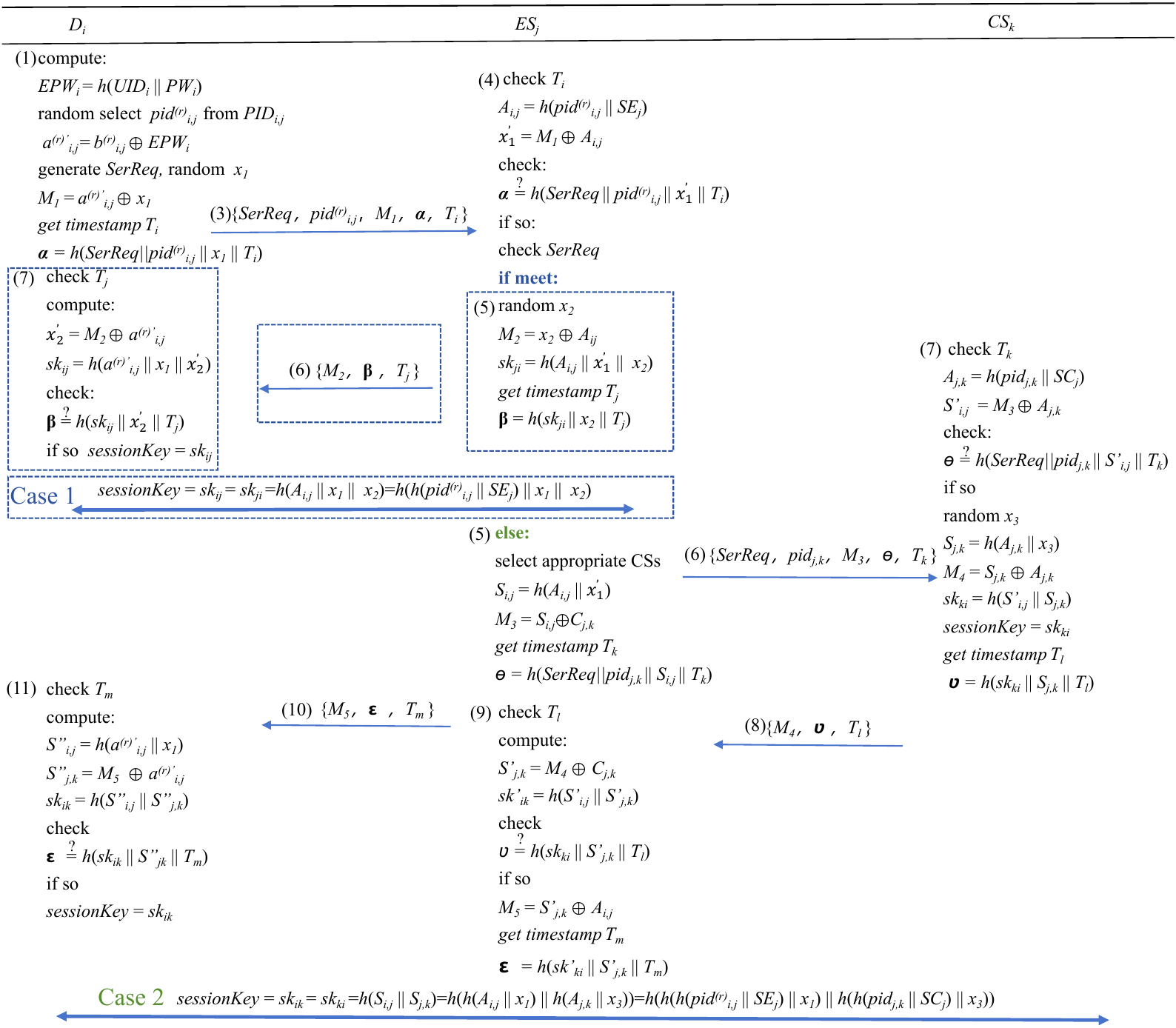}
	\caption{Authentication and key agreement.}
	\label{fig_1}
\end{figure*}
\begin{itemize}
	\item[1)] When $D_i$ initiates an authentication request to $ES_j$, $D_i$ has already received the public identifier $h(PK_j)$ broadcasted by $ES_j$ within its geographical range.
	$D_i$ computes $EPW_i = h(UID_i||PW_i)$. $D_i$ randomly selects $pid^{(r)}_{i,j}$ from $PID_{i,j}$ to authenticate with $ES_j$. It then computes $a^{(r)'}_{i,j} = EPW_i \oplus b^{(r)}_{i,j}$. Subsequently, $D_i$ generates a random value $x_1$, calculates $M_1 = a^{(r)'}_{i,j} \oplus x_1$, and computes the hash value $\alpha = h(SerReq||pid^{(r)}_{i,j}||x_1||T_i)$, where $T_i$ is timestamp. $D_i$ then sends $\{SerReq, pid^{(r)}_{i,j},M_1,\alpha,T_i\}$ to $ES_j$, where $SerReq$ represents the relevant information of the requested service.
	
	\item[2)] Upon receiving the authentication request message $\{SerReq, pid^{(r)}_{i,j},M_1,\alpha,T_i\}$ from $D_i$, $ES_j$ performs a timestamp $T_i$ verification. If $T_i$ has expired, $ES_j$ rejects the authentication request. Otherwise, $ES_j$ proceeds with the following steps: $ES_j$ retrieves $D_i$'s authentication pseudonym $pid^{(r)}_{i,j}$ from the message and computes $A_{i,j} = h(pid^{(r)}_{i,j} || SE_j)$ using its own authentication credential. Then, $ES_j$ calculates $x_1' = A_{i,j} \oplus M_1$. Futhermore, $ES_j$ computes the hash value $\alpha' = h(SerReq|| pid^{(r)}_{i,j} || x_1' || T_i)$. $ES_j$ compares the computed hash value $\alpha'$ with the hash value $\alpha$ received from $D_i$. If they are not identical, $D_i$'s authentication fails. Otherwise, $D_i$'s authentication is successful.
	If authentication is successful, $ES_j$ will checks $SerReq$. If it does not meet the requirements, it proceeds to Case 2. If it does meet the requirements, it proceeds by randomly generating $x_2$ and computing $M_2 = A_{i,j} \oplus x_2$. $ES_j$ also generates a session key $sk_{ji} = h(A_{i,j} || x'_1 || x_2)$. Finally, $ES_j$ extracts the current timestamp $T_j$, computes the hash value $\beta = h(sk_{ji} || x_2 || T_j)$, and sends the message $\{M_2, \beta, T_j\}$ back to $D_i$.
	
	\item[3)]Upon receiving the authentication response message $\{M_2, \beta, T_j\}$ from $ES_j$, $D_i$ performs a timestamp $T_j$ verification. If the timestamp $T_j$ has expired, $D_i$ rejects the response. Otherwise, $D_i$ proceeds with the following steps:$D_i$ calculates $x_2' = M_2 \oplus a^{(x)'}_{i,j}$ and $sk_{ij} = h(a^{(x)'}_{i,j} || x_1 || x'_2)$. Then, $D_i$ calculates the hash value $\beta' = h(sk_{ij} || x_2' || T_j)$. Finally, $D_i$ compares the computed hash value $\beta'$ with the hash value $\beta$ received from $ES_j$. If they are not identical, $ES_j$'s authentication fails. Otherwise, authentication and key agreement between $D_i$ and $ES_i$ have been successfully completed. $D_i$ stores the session key $sessionKey = sk_{ij}$ for future communication with $ES_j$.
\end{itemize}

$\bullet$ {\bf{\textit{Case 2}}}:

This scenario occurs when $ES_j$'s services are unable to fulfill the user's demands, necessitating $ES_j$ to request cloud services on behalf of $D_i$. The authentication method is equally applicable in a multi-cloud environment. Due to the diverse nature of user service demands, $ES_j$ can simultaneously select multiple cloud services for the user, transparently facilitating authentication between the user and CSs.

\begin{itemize}
	\item[1)] 
	The process remains the same as in Case 1 step 1).
	
	\item[2)] Upon receipt of the authentication request message $\{SerReq,pid^{(x)}_{i,j}, M_1, \alpha, T_i\}$ from $D_i$, $ES_j$ conducts a timestamp verification on $T_i$. Should $T_i$ be expired, $ES_j$ rejects the authentication request. Otherwise, $ES_j$ proceeds with the subsequent steps: retrieving $D_i$'s authentication pseudonym $pid^{(x)}_{i,j}$ from the message and computing $A_{i,j} = h(pid^{(x)}_{i,j} || SE_j)$ utilizing its own authentication credential. Subsequently, $ES_j$ calculates $x_1' = A_{i,j} \oplus M_1$. Furthermore, $ES_j$ computes the hash value $\alpha' = h(SerReq|| pid^{(x)}_{i,j} || x_1' || T_i)$. $ES_j$ compares the computed hash value $\alpha'$ with the hash value $\alpha$ received from $D_i$. If they are not identical, $D_i$'s authentication fails. Conversely, $D_i$'s authentication succeeds. Then $ES_j$ computes authentication credential $S_{i,j} = h(A_{i,j} || x_1')$. Then, based on $D_i$'s service requirements, $ES_j$ selects an appropriate cloud service provider $CS_k$ and queries the authentication pseudonym $pid_{j,k}$ and credential $C_{j,k}$ associated with the chosen $CS_k$. Next, $ES_j$ calculates $M_3 = S_{i,j} \oplus C_{j,k}$. After that, $ES_j$ extracts the current timestamp $T_k$, computes the hash value $\theta = h(SerReq||pid_{j,k} || S_{i,j} || T_k)$, and finally sends the message $\{SerReq,pid_{j,k},M_3,\theta,T_k\}$ to $CS_k$.
	
	\item[3)]Upon receiving the authentication request message $\{SerReq,pid_{j,k},M_3,\theta,T_k\}$ from $ES_j$, $CS_k$ performs a timestamp $T_k$ verification. If the timestamp $T_k$ has expired, $CS_k$ rejects the authentication request. Otherwise, $CS_k$ proceeds with the following steps: $CS_k$ retrieves $ES_j$'s pseudonym $pid_{j,k}$ from the authentication message and computes $A_{j,k} = h(pid_{j,k} || SC_k)$. Then, $CS_k$ calculates $D_i$'s authentication credential $S_{i,j}' = M_3 \oplus A_{j,k}$. Next, $CS_k$ computes the hash value $\theta' = h(SerReq||pid_{j,k} || S_{i,j}' || T_k)$. Futhermore, $CS_k$ compares the computed hash value $\theta'$ with the hash value $\theta$ received from $ES_j$. If they are not identical, $ES_j$'s authentication fails. Otherwise, $CS_k$ randomly generates $x_3$, computes the authentication credential $S_{j,k} = h(A_{j,k} || x_3)$, and then calculates $M_4 = S_{j,k} \oplus A_{i,j}$ and generates a session key $sk_{ki} = h(S_{i,j}' || S_{j,k})$. Finally, $CS_k$ extracts the current timestamp $T_l$, computes the hash value $\nu = h(sk_{ki}, S_{j,k}, T_l)$, and sends the message $\{M_4,\nu,T_l\}$ back to $ES_j$.
	
	\item[3)] Upon receiving the authentication response message $\{M_4,\nu,T_l\}$ from $CS_k$, $ES_j$ performs a timestamp $T_l$ verification. If the timestamp $T_l$ has expired, $ES_i$ rejects the response. Otherwise, $ES_j$ proceeds with the following steps: $ES_j$ compute the authentication credential $S_{j,k}' = M_4 \oplus C_{j,k}$, then calculate the key $sk'_{ki} = h(S_{i,j}' || S_{j,k}')$. Finally, $ES_j$ computes the computed hash value $\nu' = h(sk'_{ki} || S_{j,k}' || T_l)$, and compare $\nu$ with the received hash value $\nu'$. If they are not identical, $CS_k$ authentication fails. Otherwise, $CS_k$ authentication is successful, and then calculate $M_5 = S_{j,k}' \oplus A_{i,j}$. Finally, $ES_j$ extracts the current timestamp $T_m$, computes the hash value $\varepsilon = h(sk'_{ki}, S'_{j,k}, T_m)$, and sends the message $\{M_5,\varepsilon,T_m\}$ back to $D_i$.

	\item[4)] Upon receiving the authentication response message $\{M_5,\varepsilon,T_m\}$ from $ES_j$, $D_i$ performs a timestamp $T_m$ verification. If the timestamp $T_m$ has expired, $D_i$ rejects the response. Otherwise, $D_i$ proceeds with the following steps: $D_i$ compute the authentication credential $S_{i,j}'' = h(a^{(r)'}_{i,j} || x_1)$, retrieve the $CS_k$'s authentication credential $S_{j,k}'' = M_5 \oplus a'^{(r)}_{i,j}$, then calculate the key $sk_{ik} = h(S_{ij}'' || S_{j,k}'')$. Finally, $D_i$ computes the computed hash value $\varepsilon' = h(sk_{ik} || S_{j,k}' || T_m)$, and compare $\varepsilon$ with the received hash value $\varepsilon'$. If they are not identical, $CS_k$ and $ES_j$ authentication fails. Otherwise, $CS_k$ and $ES_j$ authentication is successful, and the session key $sessionKey = sk_{ik}$ is stored.
	
\end{itemize}

\subsection{Password Update}

As is well-known, passwords are susceptible to offline password guessing attacks. To enhance security, users can periodically change their passwords.
\begin{itemize}
	
\item[1)] 
First, the user inputs $\{UID_i, ID_i, PW_i\}$ to $D_i$.
\item[2)] 
$D_i$ computes $Q'_i = h(UID_i||ID_i||PW_i)$ and checks if $Q$ is identical to $Q'$. If they are not the same, the password modification fails. Otherwise, $D_i$ computes $EPW_i = h(UID_i||PW_i)$, and the user inputs the new password $PW_i'$. Then, $D_i$ calculates $EPW' = h(UID_i||PW_i')$. The authentication credential is updated as $b^{(x)'}_{i,j} = EPW_i \oplus b^{(x)}_{i,j} \oplus EPW_i'$, and the new $Q' = h(UID_i||ID_i||PW_i')$. Finally, $D_i$ saves $(Q_i', DID_i, SK_i, (h(PK_j), PID_{i,j}, B_{i,j}'))$. With this, the password update process is complete.

\end{itemize}

\section{SECURITY PROOF AND ANALYSIS}

This section primarily analyzes and proves the security of AEAKA. We first conduct a formal security analysis of the proposed scheme using the ROR model\cite{cui2019extensible,zhang2020smaka,xie2024effectively}\cite{srinivas2018anonymous,abdalla2005password,chang2015provably,dua2017secure,vinoth2020secure,guo2021anonymous}. Additionally, we provide an informal security analysis, outlining some of the known attacks that AEAKA can resist.

\subsection{Security Model}

Before formally analyzing the session key security of AEAKA, let's introduce the security primitives under the ROR model:
\begin{itemize}
	
	\item[$\bullet$] {\bf{Participants.}} The authentication process mainly involves three types of participants, including Device, ES, and CS. Each participant can execute various instances, also known as oracles. Let $D^a_i$, $ES^b_j$, and $CS^c_k$ represent the instances $a$, $b$, and $c$ of $D_i$, $ES_j$ and $CS_k$, respectively.
	
	\item[$\bullet$] {\bf{Accepted State.}} An instance $D^a_i$ transitions to an accepted state upon receiving the final expected protocol message. The session identifier ($sid$) of $D^a_i$ for the ongoing session is formed by concatenating all messages exchanged (both received and sent) by $D^a_i$ in chronological order.
	
	\item[$\bullet$] {\bf{Partnering.}} If we define the instances $D^a_i$ and $CS^c_k$ as partners, they must meet the following criteria: 1) Both $D^a_i$ and $CS^c_k$ are in the accepted state; 2) $D^a_i$ and $CS^c_k$ authenticate each other and possess the same $sid$; 3) $D^a_i$ and $CS^c_k$ are reciprocally designated as partners. Similarly, $D^a_i$ and $ES^b_j$ follow the same conditions as described above.
	
	\item[$\bullet$] {\bf{Freshness.}} An instance $D^a_i$ or $CS^c_k$ is considered fresh if the session key $sk_{ik}$ or $sk_{ki}$ between device $D^a_i$ and CS $CS^c_k$ has not been disclosed to the adversary $\mathcal{A}$. Similarly, $D^a_i$ and $ES^b_j$ follow the same conditions as described above.
	
\end{itemize}

Assuming adversary $\mathcal{A}$ has full control over all communications within the Cloud-Edge-Device network. Additionally, $\mathcal{A}$ has the capability to intercept and modify all exchanged messages and can inject new messages into the network. $\mathcal{A}$ is capable of executing the following Oracle queries:

\begin{itemize}
	
	\item[$\bullet$] \textit{Hash(m)}: Upon receiving a hash query with message $m$ from adversary $\mathcal{A}$, $\mathcal{C}$ checks if the tuple ($m, h(m)$) exists in the hash list. If it does, $\mathcal{C}$ returns $h(m)$; otherwise, $\mathcal{C}$ randomly selects $r$ and assigns $h(m) = r$. After returning $r$ to $\mathcal{A}$, $\mathcal{C}$ adds the tuple ($m, h(m)$) to the hash list.
	
	\item[$\bullet$] \textit{Execute($D^a_i, ES^b_j, CS^c_k$)}: This query simulates a passive eavesdropping attack. $\mathcal{A}$ executes this query to obtain all messages exchanged between the honest participants $D_i$, $ES_j$, and $CS_k$.
	
	\item[$\bullet$] \textit{Send($D^a_i, ES^b_j, CS^c_k$, Msg)}: This query simulates an active attack initiated by $\mathcal{A}$, such as a replay attack. $\mathcal{A}$ sends message $Msg$ to the participant instances $D^a_i$, $ES^b_j$, or $CS^c_k$ and receives the response message from these instances.
	
	\item[$\bullet$] \textit{Reveal($D^a_i, ES^b_j, CS^c_k$)}: This query simulates $\mathcal{A}$'s ability to obtain session keys. Through this query, $\mathcal{A}$ can obtain the temporary session key $sk_{ij}$ or $sk_{ik}$ created by instance $D^a_i$ and its partner $ES^b_j$, or $CS^c_k$.
	
	\item[$\bullet$] \textit{CorruptDevice($D^a_i$)}: This query simulates an attack where $\mathcal{A}$ steals information stored in a device, obtaining sensitive data.
	
	\item[$\bullet$] {\textit{Test($D^a_i, ES^b_j, CS^c_k$)}}: This query models the semantic security of session keys. When $\mathcal{A}$ executes this query on an instance $D^a_i$ ( or $ES^b_j, CS^c_k$),  where a session key has not been established, it returns the symbol $\bot$. If the session key of instance $D^a_i$ ( or $ES^b_j, CS^c_k$) has been established and is fresh, the $Test$ query randomly selects a bit $b \in \{0, 1\}.$ If $b$ = 1, it returns the actual session key; if $b$ = 0, it returns a randomly generated string of the same length as the session key. $\mathcal{A}$'s objective is to guess the value of the random bit $b$. If $\mathcal{A}$ correctly guesses $b$ with non-negligible probability, it is considered to have successfully compromised the semantic security of the session key.
	
\end{itemize}

\subsection{Formal Security Proof}

We employ the game sequence method  to provide a formal security proof for the proposed authentication and key agreement scheme(AKA). First, we list the following definition provided by \cite{guo2021anonymous,xie2024effectively}:

{\textit{Definition 1 (Semantic Security of Session Keys)}}: In the ROR model, an adversary $\mathcal{A}$ needs to distinguish between the true session key of an instance and a random key. $\mathcal{A}$ can execute $Hash$, $Execute$, $Send$, $Reveal$, $Corrupt$, and $Test$ queries on instance $D^a_i$ ( or $ES^b_j, CS^c_k$). At the end of the game, $\mathcal{A}$ guesses the value $b'$ of $b$ in the $Test$ query, and if $b = b'$, then he/she wins the game. Let $Succ$ be the event that $\mathcal{A}$ wins the game. The advantage of $\mathcal{A}$ in breaking the semantic security of the session keys of the proposed scheme $\mathcal{P}$ is defined as $Adv^{AKA}_{\mathcal{P}}(\mathcal{A})$ = $|$2$Pr[Succ] - 1$$|$. For any probabilistic polynomial-time adversary $\mathcal{A}$, if there exists a negligible function $\varepsilon$ such that $Adv^{AKA}_{\mathcal{P}}(\mathcal{A}) \le \varepsilon$, we say that the proposed scheme $\mathcal{A}$ is semantically secure in the ROR model.

{\textit{Theorem 1}}: Suppose there exists an adversary $\mathcal{A}$ aiming to compromise the semantic security of the proposed scheme $\mathcal{P}$ and derive session keys between device, ES, and CS within polynomial time. Then, the advantage of $\mathcal{A}$ in this regard is given by:

\begin{equation*}
	Adv^{AKA}_{\mathcal{P}}(\mathcal{A}) \le \frac{q^2_h}{2^{l+1}} + q_s \cdot \max \{\frac{1}  {|D|}, \frac{1}  {2^{l}}\}
\end{equation*}

Here, $q_h$, $q_s$, and $l$ respectively represent the number of Hash queries, the number of Send queries, and the length of hash values.

{\textit{Proof}}: We define four games $Game_i$, ($i$ = 0, 1, 2, 3) to demonstrate the security of the session key agreement in the protocol. Let $Succ_i$, represent the event in which the random bit $b$ in $Game_i$ is successfully guessed by $\mathcal{A}$.

{\textit{$Game_0$}}: This game simulates the original attack. In this game, $\mathcal{C}$ simulates the oracle queries like a real player. Therefore, the probability of success in this experiment is equal to the probability of adversary $\mathcal{A}$ successfully attacking the real protocol. According to the definition of semantic security \cite{abdalla2005password,chang2015provably,xie2024effectively}, we can obtain 
\begin{equation}
	Adv^{AKA}_{\mathcal{P}}(\mathcal{A}) = |2Pr[Succ_0] - 1|
\end{equation}

\textit{$Game_1$}: This game models an eavesdropping scenario orchestrated by $\mathcal{A}$. Within this game, $\mathcal{A}$ is empowered to execute multiple \textit{Execute($D^a_i, ES^b_j, CS^c_k$)} queries, capturing all message exchanges. At the game's end, $\mathcal{A}$ interrogates the system with a \textit{Test($D^a_i, ES^b_j, CS^c_k$)} query to ascertain the authenticity of the revealed session key.

In this context, $\mathcal{A}$ attempts to derive the session key $sk_{ij}$ or $sk_{ik}$, necessitating computations of $sk_{ji} = h(A_{i,j} || x_1 || x_2)$ or $sk_{ki} = h(h(A_{i,j} || x_1) || A_{j,k} || x_3)$. However, given the secrecy surrounding parameters like $x_1$, $x_2$, $x_3$, $A_{i,j}$, and $A_{j,k}$, $\mathcal{A}$'s eavesdropping endeavors in $Game_1$ fail to enhance the likelihood of triumphing in the indistinguishability game. Thus, we have 
\begin{equation}
	Pr[Succ_1] - Pr[Succ_0] = 0
\end{equation}

\textit{$Game_2$}: This game extends upon $Game_1$ by introducing $Send$ and $Hash$ queries, enabling a more sophisticated attack by $\mathcal{A}$. $Game_2$ emulates an active attack orchestrated by $\mathcal{A}$, with the objective of crafting a fraudulent message to deceive the recipient into believing it is legitimate.

Within this game, $\mathcal{A}$ is empowered to execute multiple $Hash$ queries to probe for potential hash collisions. Given that each message in the protocol comprises random numbers, timestamps, and secret parameters, the likelihood of collision during a $Send$ query executed by $\mathcal{A}$ can be disregarded.

Regarding $Hash$ queries, the probability of collision is estimated at approximately ${q^2_h} / 2^{l+1}$ according to the birthday paradox. Thus, we have

\begin{equation}
	Pr[Succ_2] - Pr[Succ_1] \le \frac{q^2_h}{2^{l+1}}
\end{equation}

\textit{$Game_3$}: This game extends $Game_2$ by incorporating the simulation of Corrupt queries. In this game, $\mathcal{A}$ executes a \textit{CorruptDevice($D^a_i$)} query to extract information stored in the device, denoted as $\{Q_i, DID_i, SK_i, (h(PK_j), PID_{i,j}, B_{i,j})\}$. Upon obtaining this information, $\mathcal{A}$ attempts to guess secret values, such as $PW_i$, $UID_i$, and $a^{(x)}_{i,j}$. Similar to the analysis in \cite{cui2019extensible}, it is challenging for $\mathcal{A}$ to deduce secret values $PW_i$ and $UID_i$ from the stored information. Let $|D|$ denote the uniform distribution password dictionary used. $\mathcal{A}$ executes $q_s$ times to guess $PW_i$, and the chance for right guessing $PW_i$ is ${q_s} / |D|$. Futhermore, $\mathcal{A}$ could guess $UID_i$ with $q_s$ times $Send$ queries, and the probability is $q_s / 2^{l}$. Obviously, the maximum probability of above is $q_s \cdot \max \{1 / |D|, 1 / 2^{l} \}$.

If the system limits the number of incorrect $UID$ and $PW$ inputs, we can derive 

\begin{equation}
	Pr[Succ_3] - Pr[Succ_2] \le q_s \cdot \max \{\frac{1}  {|D|}, \frac{1}  {2^{l}} \}
\end{equation}

All queries are simulated, and after executing the $Test$ query, $\mathcal{A}$ only needs to guess the random bit $b$ to win the game. Clearly, the probability of winning $Game_3$  is equal to the probability of guessing the random bit value, thus 

\begin{equation}
	Pr[Succ_3] = \frac{1}{2}
\end{equation}

Based Equations (2)-(4), we have

\begin{equation}
	\begin{split}
\mid Pr[Succ_3] - Pr[Succ_0]\mid    & \\
	=\mid Pr&[Succ_3] - Pr[Succ_2]\mid \\
	+\mid Pr&[Succ_2] - Pr[Succ_1]\mid \\
	+\mid Pr&[Succ_1] - Pr[Succ_0]\mid \\
	\le  \frac{q^2_h}{2^{l+1}}& + q_s \cdot \max \{\frac{1}  {|D|}, \frac{1}  {2^{l}}\}
	\end{split}
\end{equation}

Combining Equation (5) and Equation (6), we can get

\begin{equation}
	\begin{split}
		\mid Pr[Succ_0] - \frac{1}{2}\mid \le  \frac{q^2_h}{2^{l+1}}& + q_s \cdot \max \{\frac{1}  {|D|}, \frac{1}  {2^{l}}\}
	\end{split}
\end{equation}

Finally, based on Equation (7) and Equation(1), we have

\begin{equation}
	\begin{split}
		Adv^{AKA}_{\mathcal{P}}(\mathcal{A}) &= \mid 2Pr[Succ_0] - 1\mid \\
		&= 2\mid Pr[Succ_0] - 1\mid \\
		&\le  \frac{q^2_h}{2^{l+1}} + q_s \cdot \max \{\frac{1}  {|D|}, \frac{1}  {2^{l}}\}
	\end{split}
\end{equation}

\subsection{Informal Security Analysis}

In this subsection, we provide informal analysis to demonstrate the security of AEAKA.

1)\textit{Anonymity}: In AEAKA, the TA computes the anonymous identity $pid^{(x)}_{i,j} = h(DID_i || h(PK_j) || T_x)$ and $pid_{j,k} = h(EID_j || h(PK_k))$ for devices and ESs. Since a one-way hash function is utilized, attackers cannot infer the real identities $UID_i, ID_i ,EID_i$ of users, devices and ESs from $pid^{(x)}_{i,j}$ and $pid_{j,k}$, nor can they deduce the unique communication identifiers $DID_i$ associated with the $UID_i$ and $ID_i$.

2)\textit{Traceability}: The anonymous identities $DID_i$ for devices and users are computed by TA as $DID_i = h(UID_i || ID_i || s)$ and $pid^{(x)}_{i,j} = h(DID_i || h(PK_j) || T_x)$. Only TA knows the correspondence between $DID_i$ and $pid^{(x)}_{i,j}$, enabling TA to trace malicious users' corresponding identity information. Thus, the scheme can track malicious users and devices while preserving the anonymity of users and devices, satisfying the property of traceability.

3)\textit{Unlinkability}: Since AEAKA employs random numbers and timestamps, such as $x_1$, $x_2$, $x_3$, $T_i$, $T_j$, $T_k$, $T_l$, and $T_m$, the messages transmitted over the network are distinct. Moreover, the $pid^{(x)}_{i,j}$ used by devices and users is random, making it impossible for attackers to distinguish whether two different messages originate from the same sender.

4)\textit{Adaptable Mutual Authentication}:
AEAKA features a highly adaptive authentication architecture that dynamically initiates authentication methods based on device needs, and devices, ESs, and CSs authenticate each other mutually. 
In Case 1, after $D_i$ sends $\alpha = h(SerReq||pid^{(r)}_{i,j} || x_1 || T_i)$ to $ES_j$, $ES_j$ must compute $A_{i,j} = h(pid^{(r)}_{i,j} || SE_j)$ to process the data sent by $D_i$, where $SE_j$ is $ES_j$'s unique secret value. Then, $ES_j$ verifies $\alpha' = h(SerReq|| pid^{(r)}_{i,j} || x_1'|| T_i)$, achieving one-way authentication of $D_i$. Finally, $ES_j$ uses its credential $A_{i,j}$ for device $D_i$ to generate $\beta = h(h(A_{i,j} || x_1'|| x_2) || x_2 || T_j)$ and sends it to $D_i$. $D_i$ verifies the value of $\beta$, achieving authentication of $ES_j$.

In Case 2, similar to Case 1, $ES_j$ first verifies $D_i$'s identity, then computes $M_3 = S_{i,j} \oplus C_{j,k}$ and $\theta = h(SerReq||pid_{j,k} || S_{i,j} || T_k)$ using its own secret value $C_{j,k}$ for $CS_k$, and sends this message to $CS_k$ to securely transmit the authentication credential $S_{i,j}$ for $D_i$ and $ES_j$. Upon receiving the message, $CS_k$ computes $A_{j,k} = h(pid_{j,k} || SC_k)$ using its own secret value $SC_k$, derives $S_{i,j}$ from $A_{j,k}$, and finally computes the session key $sk_{ki} = h(S_{i,j}' || S_{j,k})$ and $\nu = h(sk_{ki} || S_{j,k} || T_l)$, sending them to $ES_j$. Then, $ES_j$ verifies the correctness of $\nu$, and sends $\varepsilon = h(sk'_{ki} || S'_{j,k} || T_m)$ to $D_i$. Finally, $D_i$ verifies the correctness of $\varepsilon$, achieving mutual authentication of $D_i$ and $CS_k$.

5)\textit{Session Key Agreement}:
In both two cases, session key agreement is achieved. In Case 1, key agreement between $D_i$ and $ES_j$ occurs, resulting in $sk_{ji} = h(A_{i,j} || x_1' || x_2)$ and $sk_{ij} = h(a^{(r)'}_{i,j} || x_1 || {x_2'})$. Similarly, in Case 2, key agreement between $D_i$ and $CS_k$ takes place, resulting in $sk_{ki} = h(S_{i,j}' || S_{j,k})$ and $sk_{ik} = h(S_{ij}'' || S''_{j,k})$.

6)\textit{Perfect Forward Secrecy}: In AEAKA, the keys are computed as follows: $sk_{ji} = h(A_{i,j} || x_1' || x_2)$ and $sk_{ki} = h(S_{i,j}' || S_{j,k}) = h(h(A_{i,j} || x_1) || h(A_{j,k} || x_3))$. Even if an attacker obtains some secret values of $D_i$, $ES_j$, and $CS_k$, such as $A_{i,j}$, the attacker cannot access the randomly generated values $x_1$, $x_2$, and $x_3$ for each entity. Thus, the proposed scheme achieves forward secrecy.

7)\textit{Resistance Against Impersonation Attack}: The proposed scheme is resilient against three types of impersonation attacks:

\begin{itemize}
	
	\item[$\bullet$] Impersonation of Device $D_i$. The attacker attempts to gather information from $\{SerReq,pid^{(r)}_{i,j}, M_1, \alpha, T_i\}$ to forge a legitimate authentication message. However, the attacker cannot obtain $x_1$ from the above message because $a^{(r)}_{i,j}$ is the unique secret value of the genuine $D_i$, making it impossible for the attacker to compute a legitimate $\alpha = h(pid^{(r)}_{i,j} || x_1 || T_i)$.
	
	\item[$\bullet$] Impersonation of $ES_j$. In Case 1, the attacker tries to gather information from $\{M_2, \beta, T_j\}$ to forge a legitimate authentication message. However, $\beta = h(sk_{ji} || x_2 || T_j) = h(h(A_{i,j} || x_1'|| x_2) || x_2 || T_j)$, and the attacker cannot obtain $A_{i,j}$ to generate a legitimate $\beta$.
	In Case 2, the attacker attempts to gather information from $\{pid_{j,k}, M_3, \theta, T_k\}$ to forge a legitimate authentication message. However, $\theta = h(pid_{j,k} || S_{i,j} || T_k) = h(pid_{j,k} || h(A_{i,j} || x_1') || T_k)$, and the attacker cannot obtain $A_{i,j}$, making it impossible to generate a legitimate $\theta$.
	
	\item[$\bullet$] Impersonation of $CS_k$. The attacker tries to gather information from ${M_4, \nu, T_l}$ to forge a legitimate authentication message. However, $\nu = h(sk_{ki} || S_{j,k} || T_l)$ = $h(h(h(A_{i,j} || x_1) || h(A_{j,k} || x_3)) || S_{j,k} || T_l)$, and the attacker cannot obtain $A_{i,j}$, $A_{j,k}$, or $x_3$, preventing the generation of a legitimate $\nu$.
	
\end{itemize}

8)\textit{Resistance Against Replay Attack}: We assume that the attacker can monitor communication between various entities. Although the attacker can intercept the messages, they contain timestamps with short lifespans and strong randomness, such as $x_1$, $x_2$, $x_3$, $T_i$, $T_j$, $T_k$, $T_l$, and $T_m$. Therefore, the proposed scheme provides a certain level of defense against replay attacks.

9)\textit{Resistance Against Man-in-the-Middle Attack}:In Case 1, during authentication step (3), in the message $\{SerReq,$$pid^{(r)}_{i,j}$, $M_1$,$ \alpha, T_i\}$ $D_i$ sends to $ES_j$ , the man-in-the-middle attacker cannot obtain the genuine value of $x_1$ or $a^{(r)'}_{i,j}$, so it cannot forge a valid $\alpha = h(SerReq||pid^{(r)}_{i,j} || x_1 || T_i)$, and the $ES_j$ authentication cannot pass. Similarly, in step (6), in the message $\{M_2, \beta, T_j\}$, the man-in-the-middle cannot obtain the genuine $x_2$, thus it cannot forge a valid $\beta = h(sk_{ji} || x_2 || T_j)$.

In Case 2, during authentication step (6), $\theta = h(SerReq||pid_{j,k} || h(A_{i,j} || x_1') || T_k)$, and the attacker cannot obtain $x_1$ and $A_{i,j}$, making it unable to forge a valid $\theta$. Similarly, in step (8) and step (10), in the messages $\{M_4, \nu, T_l\}$ and $\{M_5, \varepsilon, T_m\}$, and the attacker cannot obtain the values of $x_1$ and $x_3$, thus it cannot forge valid $\nu$ and $\varepsilon$.

10)\textit{Resistance Against Offline Password Guessing Attack}: During the registration process, the password $PW_i$ can be easily changed by legitimate users. Therefore, it is computationally infeasible for anyone to correctly guess $PW_i$ in polynomial time.

11)\textit{Resistance Against Insider Attacks}:During the registration process for users and devices, legitimate user $U_i$ submits a registration request message $\{UID_i, ID_i, EPW_i\}$ to the registered TA, where $EPW_i = h(UID_i || PW_i)$. Due to the one-way property of the hash function $h(\cdot)$, it is challenging for privileged insiders in TA to obtain $PW_i$.

12)\textit{Resistance Against Theft of Devices' Secret Data}: Suppose an attacker steals the information stored on $D_i$, such as $\{Q_i, DID_i, SK_i, (h(PK_j), PID_{i,j}, B_{i,j})\}$. However, since the attacker does not know the password $PW_i$, it also means the attacker cannot compute a valid $EPW_i = h(ID_i || PW_i)$. Therefore, they cannot calculate the core authentication credential $a^{(x)}_{i,j} = b^{(x)}_{i,j} \oplus EPW_i$, hence unable to complete the authentication process.

13)\textit{Resistance against Semi-Trusted ES and CS}: $D_i$ consistently uses the anonymous identity $pid^{(x)}_{i,j}$ during the authentication process. Apart from the trusted TA, no other system participant knows the real identity of $D_i$. Moreover, assuming ES and CS are semi-trusted and may leak $D_i$'s sensitive information, our scheme ensures that during the authentication process, $D_i$ does not provide sensitive information to ES, and similarly, ES does not provide any sensitive information to CS. This is because during registration, these three entities generate corresponding associated secret values, and during authentication, only these secret values are used to achieve the authentication purpose.

\section{PERFORMANCE ANALYSIS}

In this section, we will perform a performance evaluation and analysis of AEAKA including computational costs and communication overhead. We will also compare the performance of AEAKA with other approaches.

\subsection{Experimental Settings}

In terms of environment configuration, to simulate a lightweight scenario, ES and CS were deployed on a virtual machine with 4GB of memory, running Ubuntu 16.04 as the operating system, and equipped with an Intel(R) Xeon(R) CPU E5-2678 V3 @ 2.50GHz. The virtual machines were built using Workstation Pro 17.
To simulate IoT devices, we used individual laptops configured with a single-core CPU and 2GB of memory, which is comparable to the computational capabilities of most smartphones.
In terms of algorithm implementation, we used MiRACL library\cite{MI}. The hash function used was SHA-256, and the elliptic curve $y^2 = x^3 + ax $$+ b (mod p)$ was employed, where $p$ is 256 bits prime number. And the keys based ECC is 256 bits.  
We discuss comparisons of computation and communication overheads in the authentication and key agreement phases of the proposed scheme and other existing related schemes, such as those by  Cui \textit{et al.} \cite{cui2019extensible}, Zhang \textit{et al.} \cite{zhang2020smaka}, Xie \textit{et al.}\cite{xie2024effectively}, Wei \textit{et al.} \cite{wei2021lightweight}, Badshah \textit{et al.} \cite{badshah2022aake}, and Ren \textit{et al.} \cite{ren2023provable}.
\renewcommand\arraystretch{1.4}

\begin{table*}[!t]
	\caption{Computation Cost Comparsion(ms)\label{tab:table1}}
	\centering
	\begin{tabular}{|c|c|c|c|c|}
		\hline
		Schemes& Device/Vehicle/TE & CS/CSP/FN/Compute Server &  TA/AS/ES/RSU & Total\\
		\hline
		\cite{cui2019extensible} & $3T_m+8T_h \approx 0.902$ & $3T_m+7T_h \approx 0.898$ & $2T_m+10T_h \approx 0.62$ & $8T_m+25T_h \approx 2.42$ \\
		\hline
		\cite{zhang2020smaka} & $3T_m+9T_h+T_{e/d} \approx 0.938$ & $3T_m+7T_h \approx 0.898$ & $2T_m+10T_h+T_{e/d} \approx 0.652$ & $8T_m+26T_h+2T_{e/d} \approx 2.488$ \\
		\hline
		\cite{xie2024effectively} & $8T_h \approx 0.032$ & $2T_h+T_{e/d} \approx 0.04$ & $7T_h+T_{e/d} \approx 0.06$ & $17T_h+2T_{e/d} \approx 0.132$ \\
		\hline
		\cite{wei2021lightweight} & $2T_m+7T_h+T_{li} \approx 0.629$ & $5T_h+T_{li} \approx 0.041$ & $T_m+11T_h+2T_{li} \approx 0.376$ & $3T_m+23T_h+4T_{li} \approx 1.046$ \\
		\hline
		\cite{badshah2022aake} & $9T_h+8T_m \approx 2.356$ & $13T_h+5T_m+T_{e/d} \approx 1.534$ & $5T_h+2T_m \approx 0.6$ & $27T_h+15T_m+T_{e/d} \approx 4.49$ \\
		\hline
		\cite{ren2023provable} & $5T_h+3T_m \approx 0.89$ & $4T_h+3T_m \approx 0.886$ & - & $9T_h+6T_m \approx 1.776$ \\
		\hline
		AEAKA Case 1& $4T_h \approx 0.016$ & - & $4T_h \approx 0.016$ & $8T_h \approx 0.032$ \\
		\hline
		AEAKA Case 2& $5T_h \approx 0.02$ & $5T_h \approx 0.02$ & $7T_h \approx 0.028$ & $17T_h \approx 0.068$ \\
		\hline
	\end{tabular}
\end{table*}
\subsection{Computation Cost Analysis}
\begin{table}[h]
	\caption{Excecution time of basic operations(ms)}
	\centering  
	\begin{tabular}{ccc}
		\hline
		\textbf{Operation} & \textbf{Description} &\textbf{Execution time(ms)}\\ 
		\hline
		$T_a$ & ECC point addition & 0.024\\
		$T_m$ & ECC point multiplication & 0.29\\
		$T_h$ & SHA-256 hash function & 0.004\\
		$T_{e/d}$ & Symmetric encryption/decryption & 0.032\\
		$T_{li}$ & Lagrange interpolation operation & 0.021\\
		\hline
	\end{tabular}
\end{table}

In this subsection, we will conduct a computational cost analysis of AEAKA and compare it with other schemes. We analyze the relatively time-consuming cryptographic operations in the proposed solution, including ECC point addition, ECC scalar multiplication, SHA-256 hash, symmetric encryption and decryption, and Lagrange interpolation operationwhich are denoted by $T_{a}$/$T_{m}$/$T_{h}$/$T_{e/d}/T_{li}$.
Each simulation experiment was repeated 200 times, and take the the average as the result. Table III shows the execution time of some basic operations used in this paper.\par

In Cui \textit{et al.}'s scheme\cite{cui2019extensible}, there are three entities involved: the vehicle, the TA, and the CSP. In this scheme, the vehicle initiates the authentication request process and receives the session key agreement message from CSP, consuming a total of $3T_m+8T_h \approx 0.902$ ms. The TA participates in the authentication of both the vehicle and the CSP, consuming a total of $2T_m+10T_h \approx 0.62$ ms. Finally, the CSP engages in information exchange with the TA and the vehicle for session key agreement, consuming a total of $3T_m+7T_h \approx 0.898$ ms. Therefore, in Cui \textit{et al.}'s scheme, the total time consumed is $8T_m+25T_h \approx 2.42$ ms.

Similar to Cui \textit{et al.}'s scheme\cite{cui2019extensible}, Zhang \textit{et al.}'s scheme \cite{zhang2020smaka} also involves three entities: the vehicle, the TA, and the CSP. While this scheme introduces a login process for users with the CSP, it is not within the scope of this analysis. Apart from this addition, the information exchange process for authentication and session key agreement in Zhang \textit{et al.}'s scheme is similar to that of Cui \textit{et al.}'s scheme. The vehicle consumes $3T_m+9T_h+T_{e/d} \approx 0.938$ ms, the TA participates in the authentication of both the vehicle and the CSP, consuming $2T_m+10T_h+T_{e/d} \approx 0.652$ ms, and the CSP consumes $3T_m+7T_h \approx 0.898$ ms. In the end, all entities collectively consume $8T_m+26T_h+2T_{e/d} \approx 2.488$ ms.

In Xie \textit{et al.}'s scheme\cite{xie2024effectively}, there are three entities involved: the Terminal End (TE), the Authentication Server (AS), and the Compute Server. The TE consumes $8T_h \approx 0.032$ ms, the AS consumes $7T_h+T_{e/d} \approx 0.06$ ms, and the Compute Server consumes $2T_h+T_{e/d} \approx 0.04$ ms. Therefore, the total time consumed for all entities is approximately $17T_h+2T_{e/d} \approx 0.132$ ms.
\begin{figure}[!t]
	\centering
	\includegraphics[width=0.5\textwidth]{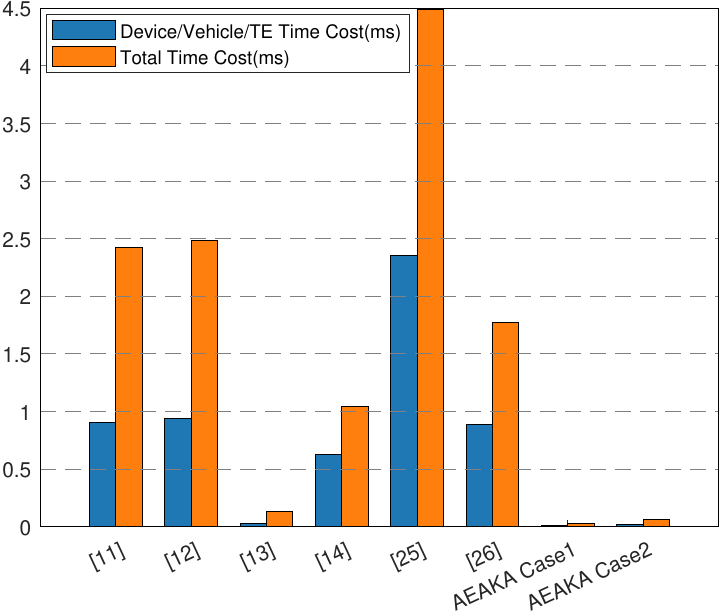}
	\caption{Computation cost comparison.}
	\label{fig_1}
	\vspace{-1.6em}
\end{figure}
In Wei \textit{et al.}'s scheme\cite{wei2021lightweight}, the involved entities are the vehicle, the Fog Node (FN), and the TA. The vehicle initiates the authentication request to the FN, which then forwards the authentication message to the TA, initiating authentication. Finally, the authentication between the vehicle and the FN is completed, involving the Lagrange algorithm. We approximate the time taken for the pseudo-random function to be approximately equal to the time taken for the hash function. Therefore, the vehicle consumes approximately $2T_m+7T_h+T_{li} \approx 0.629$ ms, the FN consumes approximately $5T_h+T_{li} \approx 0.041$ ms, and the TA consumes approximately $T_m+11T_h+2T_{li} \approx 0.376$ ms. Ultimately, Wei \textit{et al.}'s scheme consumes approximately $3T_m+23T_h+4T_{li} \approx 1.046$ ms.

In Badshah \textit{et al.}'s scheme\cite{badshah2022aake}, there are three entities involved: the vehicle, the Road Side Unit (RSU), and the Cloud Server (CS). This scheme provides authentication between the vehicle and the RSU, and between the RSU and the CS. Therefore, if the vehicle wishes to request CS services, it must first authenticate with the RSU and then with the CS through the RSU. Since the consumption of Physical Unclonable Function (PUF) is negligible, it is not considered. Consequently, the vehicle consumes approximately $9T_h+8T_m \approx 2.356$ ms, the RSU consumes approximately $5T_h+2T_m \approx 0.6$ ms, and the CS consumes approximately $13T_h+5T_m+T_{e/d} \approx 1.534$ ms. The total time consumed for all entities in the process is approximately $27T_h+15T_m+T_{e/d} \approx 4.49$ ms.

In Ren \textit{et al.}'s scheme\cite{ren2023provable}, there are two entities involved: the Device and the CS. The authentication process is relatively simple in this scheme. The Device consumes approximately $5T_h+3T_m \approx 0.89$ ms, while the CS consumes approximately $4T_h+3T_m \approx 0.886$ ms. Therefore, the total time consumed for all entities is approximately $9T_h+6T_m \approx 1.776$ ms.

In AEAKA, there are three entities involved: the Device, the ES, and the CS. For ease of comparison with other schemes, we first consider Case 1. The Device consumes approximately $4T_h \approx 0.016$ ms, and the ES consumes approximately $4T_h \approx 0.016$ ms,. Since in Case 1, ES already meets the device's requirements without needing to request CS, the total time consumption is $8T_h \approx 0.032$. Then, in Case 2, the Device consumes approximately $5T_h \approx 0.02$ ms, the ES consumes approximately $7T_h \approx 0.028$ ms, and the CS consumes approximately $5T_h \approx 0.02$ ms. Therefore, the total time consumed for all entities is approximately $17T_h \approx 0.068$ ms.

The comparative analysis of the time consumption for each scheme is presented in Table IV and Fig. 7, showing that AEAKA demonstrates significantly better computational efficiency compared to others. The time expenditure at the devices in AEAKA is notably lower than in other schemes, primarily because AEAKA's devices do not undergo complex cryptographic operations such as ECC, but rather simple hash function computations. This also indicates that AEAKA is more adaptable to a wide range of resource-constrained devices. In AEAKA, the authentication process is assisted by ES, eliminating the need for frequent interactions with TA, thereby greatly reducing TA's overhead and enhancing system security.

\subsection{Communication Cost Analysis}

In the Experimental Settings section, as mentioned, we set the length of the hash function to 256 bits, the length of the timestamp T to 32 bits, the length of the points on ECC to 256 bits, and the length of the symmetric encryption to 512 bits.

In Cui \textit{et al.}'s scheme\cite{cui2019extensible}, for the sake of fairness, we do not consider the service request message $M_i$ from the vehicle. Besides $M_i$, the scheme involves a total of 5 messages, denoted as $Msg_1$ =$ \{PID_i,X,\eta,tt_i\}$, $Msg_2 = \{PID_j,Y,\theta,tt_i\}$, $Msg_3 = \{CID_i,AID_i,\alpha,\beta,X,tt_i\}$, $Msg_4 = \{AID_j,Y,\beta,\gamma,tt_i\}$, and $Msg_5 = \{\lambda\}$. Therefore, $|Msg_1|$ = 256 + 2562 + 256 + 32 = 1056 bits, $|Msg_2|$ = 256 + 2562 + 256 + 32 = 1056 bits, $|Msg_3|$ = 256 + 256 + 256 + 256 + 2562 + 32 = 1568 bits, $|Msg_4|$ = 256 + 2562 +256 + 256 + 32 = 1312 bits, and $|Msg_5|$ = 256 bits. The total communication overhead of this scheme is calculated as: 1056$\times$2 + 1568 + 1312 = 4992 bits.

In Zhang \textit{et al.}'s scheme \cite{zhang2020smaka}, there are three message interactions, denoted as $Msg_1 = \{W_i,PID_i,F_i\}$, $Msg_2 = \{N_r,W_i'',GID_r,TS_r,\theta,\rho\}$, and $Msg_3 = \{L_j,GID_r,\rho,TS_r,TS_{ij}\}$. Therefore, $|Msg_1|$ = 2562 + 512 + 256 + 32 = 1312 bits, $|Msg_2|$ = 256 + 256 + 256 + 32 + 256 + 256 = 1312 bits, and $|Msg_3|$ = 256 + 256 + 256 + 32 + 32 = 832 bits. The total communication overhead of this scheme is calculated as: 13122 + 832 = 3456 bits.

In Xie \textit{et al.}'s scheme\cite{xie2024effectively}, there are three message interactions, denoted as $Msg_1 $=$ \{SHID_{TE},M_1,M_2,AU_{TE},T_1\}$, $Msg_2 = \{M_5\}$, and $Msg_3 = \{M_6,AU_S,T_3\}$. Therefore, $|Msg_1|$ = 256 + 256 + 256 + 256 + 32 = 1056 bits, $|Msg_2|$ = 512 bits, and $|Msg_3|$ = 256 + 256 + 32 = 544 bits. The total communication overhead of this scheme is calculated as: 1056 + 512 + 544 = 2112 bits.
\begin{table}[!t]
	\caption{Communication Cost Comparsion\label{tab:table1}}
	\centering
	\begin{tabular}{|c|c|c|}
		\hline
		Schemes&   Number of messages &  Length of message\\
		\hline
		\cite{cui2019extensible} & 5 & 4992 bits\\
		\hline
		\cite{zhang2020smaka} &  3 & 3456 bits\\
		\hline
		\cite{xie2024effectively} & 3 & 2112 bits\\
		\hline
		\cite{wei2021lightweight} & 4 & 4786 bits\\
		\hline
		\cite{badshah2022aake} & 4 & 4224 bits\\
		\hline
		\cite{ren2023provable} & 2 & 1792 bits\\
		\hline
		AEAKA Case 1& 2 & 1344 bits\\
		\hline
		AEAKA Case 2& 4 & 2688 bits\\
		\hline
	\end{tabular}
\end{table}
In Wei \textit{et al.}'s scheme\cite{wei2021lightweight}, there are four message interactions, denoted as $Msg_1= \{pid_i,t_i,R_i,\alpha_i\}$, $Msg_2=\{id_j,t_j,n_j,pid_i,t_i,R_i,\alpha_i,\beta_i\}$, $Msg_3 = \{n_k,fn_k,t_k,\gamma_k,\sigma_k\}$, and $Msg_4 = \{n_k,fn_k,t_k,\sigma_k\}$. Therefore, $|Msg_1|$ = 256 + 32 + 2562 + 256 = 1056 bits, $|Msg_2|$ = 256 + 32 + 256 + 256 + 32 + 2562 + 256 + 256 = 1856 bits, $|Msg_3|$ = 256 + 256 + 32 + 256 + 256 = 1056 bits, and $|Msg_4|$ = 256 + 256 + 32 + 256 = 800 bits. The total communication overhead of this scheme is calculated as: 1056 + 1856 + 1056 + 800 = 4786 bits.

In Badshah \textit{et al.}'s scheme\cite{badshah2022aake}, from the vehicle to CS, there are four message interactions, denoted as $Msg_1 $=$ \{M_1,Auth_1,BSC,TS_1\}$, $Msg_2 = \{M_2,Auth_3,TS_2\}$, $Msg_3 = \{Z_3,BSC,Z_4,Z_5,TS_1\}$, and $Msg_4 = \{Z_6,Z_7,TS_2\}$. Therefore, $|Msg_1|$ = 2562 + 256 + 2562 + 32 = 1312 bits, $|Msg_2|$ = 256 $\times$ 2 + 256 + 32 = 800 bits, $|Msg_3|$ = 2562 + 2562 + 256 + 256 + 32 = 1568 bits, and $|Msg_4|$ = 256 + 256 + 32 = 544 bits. The total communication overhead of this scheme is calculated as: 1312 + 800 + 1568 + 544 = 4224 bits.

\begin{figure}[!t]
	\centering
	\includegraphics[width=0.46\textwidth]{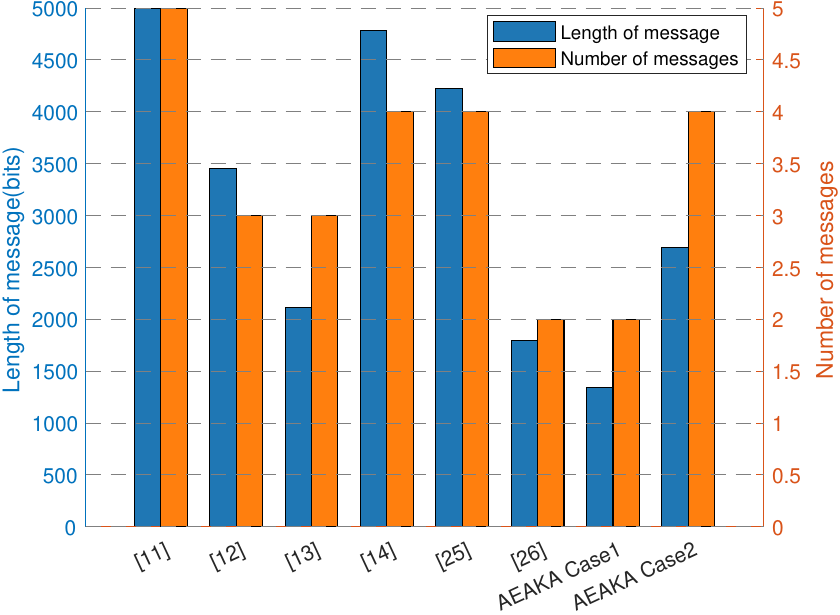}
	\caption{Communication cost comparison.}
	\label{fig_1}
	\vspace{-1.6em}
\end{figure}
\begin{table*}[h]
	\caption{Comparison of Security and Functionality Features}
	\centering  
	\begin{tabular}{ccccccccc}
		\hline
		Security and Functionality Features &\cite{cui2019extensible} &\cite{zhang2020smaka} &\cite{xie2024effectively}&  \cite{wei2021lightweight}&    \cite{badshah2022aake} & \cite{ren2023provable} & AEAKA\\ 
		\hline
		Anonymity & \Checkmark & \Checkmark &\Checkmark&\Checkmark &\XSolidBrush &\Checkmark  & \Checkmark\\
		Traceability &\Checkmark &\Checkmark &\Checkmark& \Checkmark&\XSolidBrush &\Checkmark  & \Checkmark\\
		Unlinkability &\Checkmark &\Checkmark &\Checkmark& \Checkmark& \Checkmark&\Checkmark  & \Checkmark\\
		Adaptable Mutual Authentication &\XSolidBrush &\XSolidBrush &\XSolidBrush&\XSolidBrush &\Checkmark & \XSolidBrush & \Checkmark\\
		Session Key Agreement &\Checkmark & \Checkmark&\Checkmark &\Checkmark&\Checkmark &\Checkmark  & \Checkmark\\
		Releasing the burden on the TA & \XSolidBrush&\XSolidBrush &\Checkmark&\XSolidBrush &\Checkmark &\Checkmark  & \Checkmark\\
		No complex ECC computations & \XSolidBrush&\XSolidBrush &\Checkmark&\XSolidBrush &\XSolidBrush &\XSolidBrush  & \Checkmark\\
		Perfect Forward Secrecy &\Checkmark &\Checkmark &\Checkmark&\Checkmark &\Checkmark & \Checkmark & \Checkmark\\
		Resistance Against Impersonation Attack &\Checkmark &\XSolidBrush &\Checkmark&\XSolidBrush &\Checkmark &\Checkmark  & \Checkmark\\
		Resistance Against Replay Attack &\XSolidBrush &\Checkmark &\Checkmark&\Checkmark &\Checkmark &\XSolidBrush  & \Checkmark\\
		Resistance Against Offline Password Guessing Attack & \Checkmark&\Checkmark &\XSolidBrush&\XSolidBrush & \XSolidBrush& \XSolidBrush & \Checkmark\\
		Resistance against Insider Attacks & \Checkmark&\Checkmark &\XSolidBrush &\XSolidBrush&\XSolidBrush &\Checkmark  & \Checkmark\\
		Resistance against Theft of Devices' Secret Data &\Checkmark & \Checkmark&\Checkmark&\XSolidBrush & \XSolidBrush&\Checkmark  & \Checkmark\\
		Resistance against Semi-Trusted ES and CS &\XSolidBrush &\XSolidBrush &\Checkmark&\XSolidBrush &\XSolidBrush &\Checkmark  & \Checkmark\\
		Resistance Against Man-in-the-Middle Attack &\XSolidBrush &\Checkmark &\Checkmark&\Checkmark &\Checkmark &\Checkmark  & \Checkmark\\
		\hline
	\end{tabular}
\end{table*}
In Ren \textit{et al.}'s scheme\cite{ren2023provable}, there are two message interactions, denoted as $Msg_1 = \{PID,M_1,Auth_1\}$ and $Msg_2 = \{M_2,Auth_2\}$. Therefore, $|Msg_1|$ = 256 + 2562 + 256 = 1024 bits, and $|Msg_2|$ = 2562 + 256 = 768 bits. The total communication overhead of this scheme is calculated as: 1024 + 768 = 1792 bits.

In AEAKA, similar to Cui \textit{et al.}'s scheme\cite{cui2019extensible}, we do not consider the service request message $SerReq$ from the device. In Case 1, there are two message interactions, denoted as $Msg_1 = \{pid_{i,j},M_1,\alpha,T_i\}$ and $Msg_2 = \{M_2,\beta,T_j\}$. Therefore, $|Msg_1|$ = 256 + 256 + 256 + 32 = 800 bits, and $|Msg_2|$ = 256 + 256 + 32 = 544 bits. Thus, Case 1 has a total communication of 800 + 544 = 1344 bits. In Case 2, there are four message interactions, denoted as $Msg_1 = \{pid_{i,j},M_1,\alpha,T_i\}$, $Msg_2 = \{pid_{j,k},M_3,\theta,T_k\}$, $Msg_3 = \{M_4,\nu,T_l\}$, and $Msg_4 = \{M_5,\varepsilon,T_m\}$. Therefore, $|Msg_1|$ = 256 + 256 + 256 + 32 = 800 bits, $|Msg_2|$ = 256 + 256 + 256 + 32 = 800 bits, $|Msg_3|$ = 256 + 256 + 32 = 544 bits, and $|Msg_4|$ = 256 + 256 + 32 = 544 bits. Thus, Case 2 has a total communication of 800 + 800 + 544 + 544= 2688 bits.

Table V shows the comparison results of communication overhead of each scheme. The communication cost comparison results indicate that in Case 2, AEAKA's communication cost is slightly higher than Xie \textit{et al.}'s scheme\cite{xie2024effectively} and Ren \textit{et al.}'s scheme\cite{ren2023provable}. This is because AEAKA can adaptively request cloud authentication through edge assistance, whereas their solutions are limited to single authentication scenarios and lack scalability.\par
Overall, AEAKA has lower communication costs compared to other schemes and possesses higher scalability.

\subsection{Comparison of Security and Functionality Features}
In this subsection, we will compare the security and functionality of AEAKA with other schemes. The comparison results are shown in Table VI, where the symbol \Checkmark indicates that the scheme is secure or provides that functionality. Conversely, the symbol \XSolidBrush indicates that the scheme is insecure or does not provide that functionality. It can be observed that AEAKA exhibits higher security performance and offers more functionalities.

\section{Conclusion}

In this paper, we propose an adaptive and efficient authentication and key agreement method (AEAKA) for Cloud-Edge-Device IoT environments. The AEAKA framework features a highly adaptive authentication architecture that dynamically initiates different authentication methods based on device needs. Additionally, our edge-assisted scheme reduces the load on trust authorities, and the lightweight authentication protocol employs hash-based algorithm, ensuring compatibility with resource-constrained devices. The use of associated authentication credentials, enhancing the privacy of the authentication process. Security proofs and performance analyses demonstrate that AEAKA outperforms existing methods in terms of security and authentication efficiency.

Despite the significant advancements presented in this work, there are areas for future exploration. The current approach to identity management, while effective, requires considerable storage space. Future research focuses on optimizing identity management to minimize storage requirements. Additionally, we plan to investigate the design of batch authentication protocols in cloud-edge-device scenarios to enhance the efficiency and scalability of authentication processes further. These future enhancements build on the foundation laid by AEAKA, ensuring robust and efficient authentication mechanisms for evolving IoT environments.

\bibliographystyle{ieeetr}

\bibliography{references}

\begin{IEEEbiography}
	[{\includegraphics[width=0.9in,height=1.25in,clip,keepaspectratio]{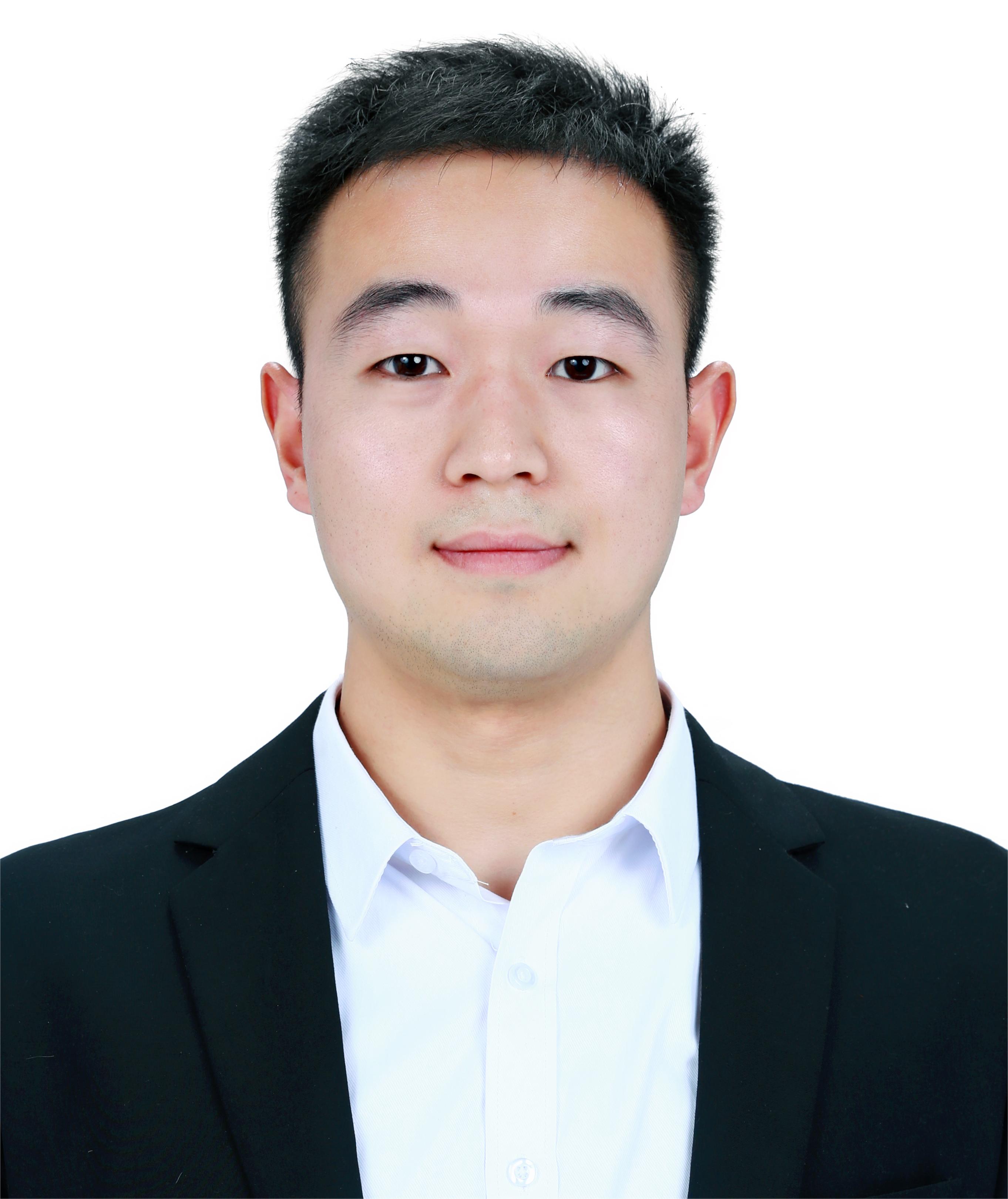}}]
	{\\ \\ \\ Kexian Liu}is currently pursuing the Ph.D. degree with the
	State Key Laboratory of Networking and Switching Technology, BUPT, Beijing, China. His research focuses on future network architecture and network security.

\end{IEEEbiography}

\begin{IEEEbiography}
	[{\includegraphics[width=1in,height=1.25in,clip,keepaspectratio]{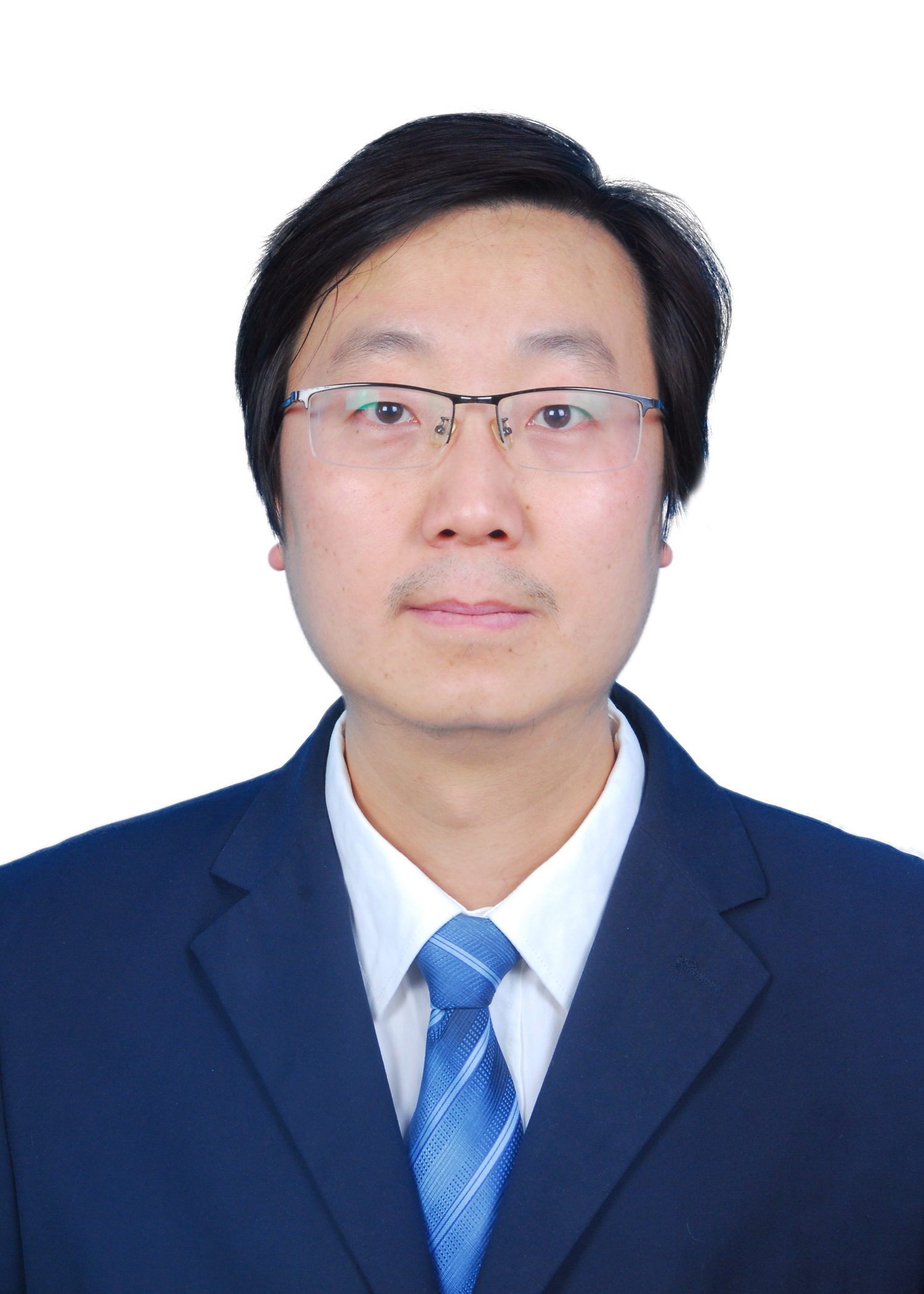}}]
	{Jianfeng Guan}(Member, IEEE) received the B.S. degree in telecommunication engineering from Northeastern University, Shenyang, Liaoning Province, China in 2004, and Ph.D degree in Communication and Information System from Beijing Jiaotong University, Beijing, China in 2010. From 2010 to 2015, he was a lecture with Institute of Network Technology, Beijing University of Posts and Telecommunications. Since 2016, he has been an Assistant Professor. He is the author of more than 100 articles and more than 70 inventions. His research interests include future network architecture, network security and mobile Internet. Dr. Guan is a recipient of several Best Paper Awards from ACM Mobility conference 2008, IC-BNMT2009, and Mobisec2018. He servers as TPC member for WCNC2019, ICC 2018 CCNCP, Globecom 2018 CCNCPS, MobiSec 2016-2019, INFOCOM MobilWorld 2011, 2015-2017, ICCE 2017. He also is a reviewer for Journals such as TVT, TB, CC, CN, JNCA, FGCS, ACCESS, INS, SCN, IJSSC, IJAHUC, JoWUA. 

\end{IEEEbiography}

\begin{IEEEbiography}
	[{\includegraphics[width=1in,height=1.25in,clip,keepaspectratio]{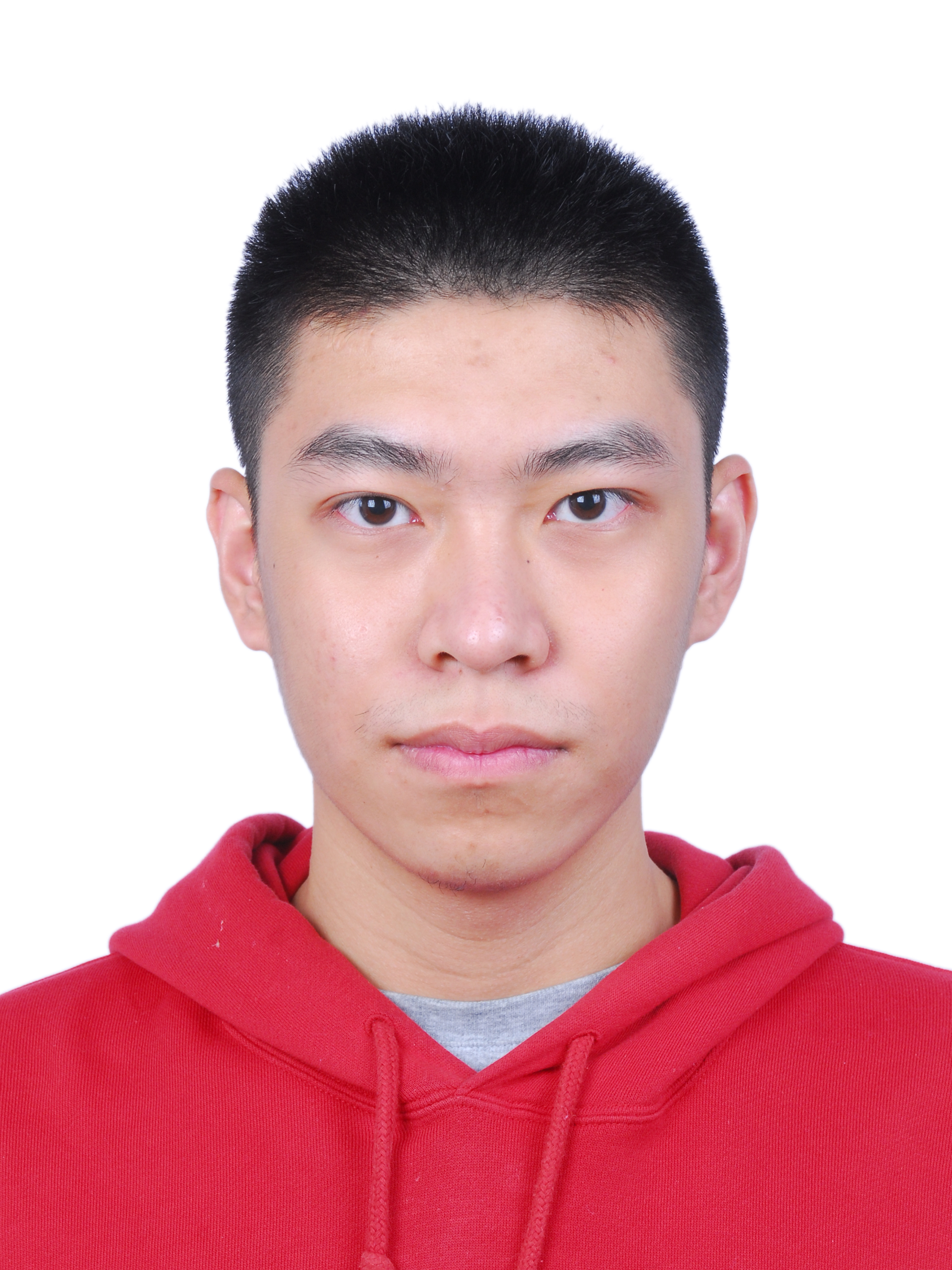}}]
	{\\ \\ \\Xiaolong Hu} is currently pursuing the master degree with the
	State Key Laboratory of Networking and Switching Technology, BUPT, Beijing, China. His research focuses on future network architecture and network security.
	
\end{IEEEbiography}
\begin{IEEEbiography}
	[{\includegraphics[width=1in,height=1.25in,clip,keepaspectratio]{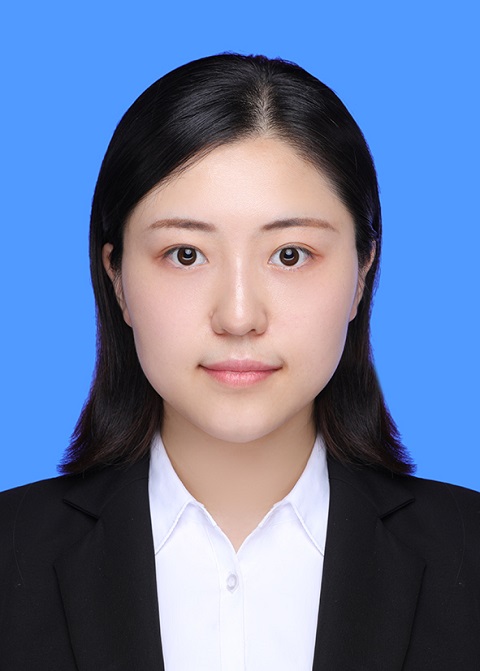}}]
	{\\ \\ \\Jing Zhang} is currently pursuing the Ph.D. degree with the
	State Key Laboratory of Networking and Switching Technology, BUPT, Beijing, China. Her research focuses on future network architecture and intelligent routing.
	
\end{IEEEbiography}

\begin{IEEEbiography}
	[{\includegraphics[width=1in,height=1.25in,clip,keepaspectratio]{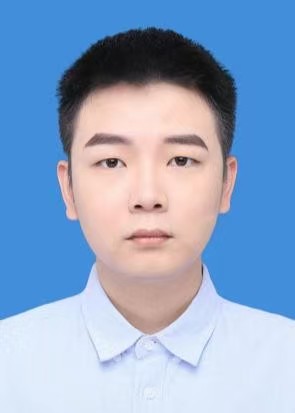}}]
	{\\ \\ \\Jianli Liu} is currently pursuing the master degree with the
	State Key Laboratory of Networking and Switching Technology, BUPT, Beijing, China. His research focuses on authenticaiton.

\end{IEEEbiography}

\begin{IEEEbiography}
	[{\includegraphics[width=1in,height=1.25in,clip,keepaspectratio]{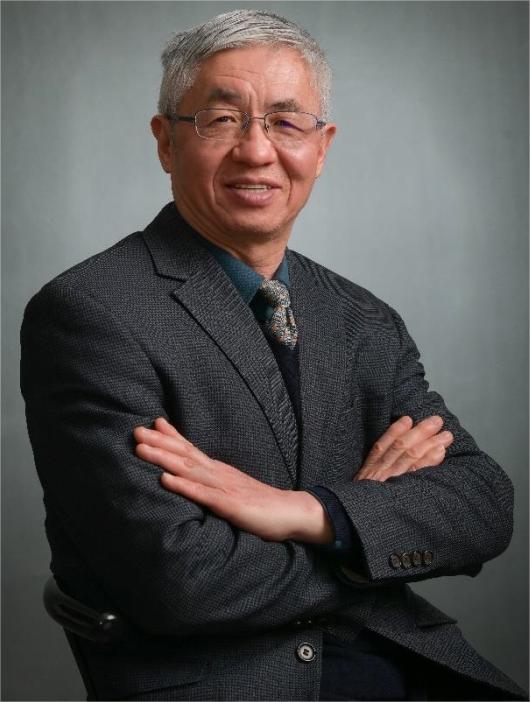}}]
	{Hongke Zhang} (Fellow, IEEE) received the Ph.D. degree in communication and information system from the University of Electronic Science and Technology of China, Chengdu, China, in 1992. He is currently a Professor with the School of Electronic and Information Engineering, Beijing Jiaotong University, Beijing, China, where he currently directs the National Engineering Center of China on Mobile Specialized Network. He is an Academician of China Engineering Academy, Beijing, and the Co-Director of the PCL Research Center of Networks and Communications, Peng Cheng Laboratory, Shenzhen China. His current research interests include architecture and protocol design for the future Internet and specialized networks. Prof. Zhang currently serves as an Associate Editor for the IEEE TRANSACTIONS ON NETWORK AND SERVICE MANAGEMENT and IEEE INTERNET OF THINGS JOURNAL.
	
\end{IEEEbiography}

\end{document}